\documentclass[numberedappendix]{emulateapj}

\newcommand{\ba}{\begin{eqnarray}}
\newcommand{\ea}{\end{eqnarray}}
\newcommand{\be}{\begin{equation}}
\newcommand{\ee}{\end{equation}}

\newcommand{\g}{\gamma}
\newcommand{\gp}{{\gamma^\prime}}

\def\e{\epsilon}
\def\ep{\epsilon^\prime}

\begin{document}

\title{Near-Equipartition Jets with Log-Parabola Electron
Energy Distribution and the Blazar Spectral-Index Diagrams
}

\author{Charles D. Dermer,\altaffilmark{1}
 Dahai Yan,\altaffilmark{2,3} Li Zhang,\altaffilmark{3}
 Justin D.\ Finke,\altaffilmark{1}
\& Benoit Lott\altaffilmark{4}
}

\altaffiltext{1}{Space Science Division, U.S. Naval
  Research Laboratory, Washington, DC 20375, USA. e-mail:
  charles.dermer@nrl.navy.mil }
\altaffiltext{2}{ Key Laboratory of Particle Astrophysics, Institute of High Energy Physics, Chinese Academy of Sciences, Beijing 100049, China}
\altaffiltext{3}{Department of Astronomy, Yunnan University, Kunming 650091, Yunnan, China}
%\altaffiltext{4}{CNRS/IN2P3, Centre d' \'Etudes Nucl\'eaires Bordeaux Gradignan, UMR 5797, Gradignan, 33175, France}
%\altaffiltext{5}{40 Universit\'e de Bordeaux, Centre d' \'Etudes Nucl\'eaires Bordeaux Gradignan, UMR 5797, Gradignan, 33175, France}
\altaffiltext{4}{Centre d'\'Etudes Nucl\'eaires Bordeaux Gradignan, Universit\'e de Bordeaux, CNRS/IN2P3, UMR 5797, Gradignan, 33175, France }

\begin{abstract}
Fermi-LAT analyses show that the $\gamma$-ray photon spectral indices
$\Gamma_\gamma$ of a large sample of blazars correlate with the $\nu F_\nu$
peak synchrotron frequency $\nu_{s}$  according to the relation
$\Gamma_\gamma = d - k \log\nu_{s}$. The same function,
with different constants $d$ and $k$, also describes the relationship between
$\Gamma_\gamma$ and peak Compton frequency $\nu_{\rm C}$.
This behavior is derived analytically using an equipartition blazar model
with a log-parabola description of the electron energy distribution (EED).
In the Thomson regime, $k = k_{EC} = 3b/4$ for external Compton processes
and $k = k_{SSC}  = 9b/16$ for synchrotron self-Compton (SSC) processes,
where $b$ is the log-parabola width parameter of the EED. The BL Lac object Mrk 501
is fit with a synchrotron/SSC model given by the log-parabola EED,
and is best fit away from equipartition. { Corrections are made to the 
spectral-index diagrams
for a low-energy power-law EED and departures
from equipartition, as constrained by absolute} jet power.
Analytic expressions are compared with numerical values derived from
self-Compton and external Compton scattered $\gamma$-ray spectra
from Ly $\alpha$ broad-line region and IR target photons.
The   $\Gamma_\gamma$ vs.\  $\nu_s$ behavior in the model depends strongly on
$b$, with progressively and predictably weaker dependences on $\gamma$-ray detection range,
variability time, and isotropic $\gamma$-ray luminosity. Implications for 
blazar unification and blazars as ultra-high energy cosmic-ray sources are presented.
{ Arguments by Ghisellini et al.\ (2014) that the jet power exceeds the accretion 
luminosity depend on the doubtful assumption that we are viewing at the 
Doppler angle.}
\end{abstract}

%OLD: gamma rays: theory---radiation mechanisms: nonthermal---galaxies: AGN: blazars
%black hole physics---astroparticle physics---gamma rays: galaxies---
%galaxies: jets---galaxies: active---radiation mechanisms: non-thermal---acceleration of particles
% (galaxies:) BL Lacertae objects: general
% (galaxies:) BL Lacertae objects: individual (..., ...)
%---galaxies: active

\section{Introduction}

Searches for an ordering principle in blazar science have met with limited  success.
One of the most debated is the {\it blazar sequence}, in which
blazar data seem to show
an inverse correlation between apparent isotropic synchrotron luminosity $L_{syn}$ and peak synchrotron frequency $\nu_s$ of the blazar $\nu F_\nu$ spectral energy distribution (SED) \citep{fos98,smu96}. This behavior, which is mirrored in the $\gamma$-ray regime, has been interpreted in terms of cooling processes \citep{ghi98,bd02,fin13}. The validity of the blazar sequence has, however, been criticized \citep{gio12,gpp13} as possibly resulting from spurious correlations introduced by combining samples from radio and X-ray blazar surveys, problems from redshift incompleteness, and confusing lineless BL Lac objects that lack accretion disk with those where the BLR radiation is overwhelmed by beamed emission.   Contrary to 
the simple blazar sequence, \citet{Meyer} present evidence  for the existence of two 
separate tracks in the $L_{syn}$ vs.\ $\nu_s$ plane,  including radio galaxies in the blazar-sequence plot. 

A second strong correlation is the {\it blazar divide}. From the first three months of Fermi Large Area Telescope (LAT) blazar data,   \cite{gmt09} argued that hard ($\Gamma_\gamma <2$) $\gamma$-ray spectrum blazars are associated with sources radiating isotropic $\gamma$-ray luminosities  $L_\gamma\lesssim 5\times 10^{46}$ erg s$^{-1}$, while soft ($\Gamma_\gamma > 2$) $\gamma$-ray blazars are more likely to be at larger values of $L_\gamma$. 
From the Second Fermi LAT AGN (2LAC) data \citep{2LAC},  a broad divide is evident in the direct data at $L_\gamma\cong 10^{46}$ erg s$^{-1}$  \citep[Figs.\ 37 and 38 in][]{2LAC}, though with no other 
apparent dependence of $\Gamma_\gamma$ on $L_\gamma$ in the ranges $10^{44} \lesssim L_\gamma\lesssim 10^{46}$ erg s$^{-1}$
and $10^{46} \lesssim L_\gamma\lesssim 10^{49}$ erg s$^{-1}$. 
In terms of a beaming-corrected Eddington ratio $\ell_{\rm Edd}$ for a black hole with mass $\sim 10^9$ M$_\odot$, this could imply a transition from an inefficiently radiating ADAF-type flow at $\ell_{\rm Edd} \lesssim 0.01$ to a thick disk when $\ell_{\rm Edd} \gtrsim 0.01$ \citep{gmt09}. 
The $\gamma$-ray Compton dominance ${\cal A}_{\rm C}$, which is essentially the ratio of the bolometric $\gamma$-ray and synchrotron luminosities, also strongly correlates with $\nu_s$ \citep{fos98,fin13}.
%Many other searches for spectral and temporal correlations and cross-correlations of blazar emissions have been made [examples: radio-$\gamma$, $\gamma$-ray phenomenology, space permitting].

Definitive interpretations of blazar sequence and blazar divide data are hampered by redshift incompleteness. BL Lac  objects without redshift information may themselves constitute separate populations in the $L_{syn}$ vs.\ $\nu_s$ or $\Gamma_\gamma$ vs.\ $L_\gamma$ planes, though large efforts have been made to provide complete, or at least redshift-constrained samples of blazar data \citep{sha13,aje14}. { The ${\cal A}_{\rm C}$ vs.\ $\nu_s$ distributions of 2LAC blazars with and without redshift do not significantly differ \citep{fin13}.}

A third robust correlation in blazar physics  relates  $\gamma$-ray spectral index $\Gamma_\gamma$ with peak synchrotron ($\nu_s$) or peak Compton $\gamma$-ray ($\nu_{\rm C}$) frequencies (in this study, we assume that
the blazar SEDs are made by leptonic processes only).\footnote{{ A correlation of the log-parabola width parameter $b$ and $\nu_s$ is apparent in SED modeling studies \citep{chen14}, but is based on only 5 or 6 high-synchrotron peaked  blazars.} } 
%The principal advantage of these {\it blazar spectral-index diagrams} over the blazar sequence and blazar divide is that they circumvent the incompletness problem, though of course such diagrams can be restricted to only those sources with known or constrained redshift.
% \citep[cf.][]{aje14}.
These {\it spectral-index diagrams} for FSRQ and BL Lac blazars have been reported in the
First LAT AGN Catalog \citep[1LAC, Fig.\ 13 in][]{1LAC}, the Fermi Bright Blazar SED paper \citep[Fig.\ 29 in][]{SED}, the 2LAC  \citep[Fig.\ 17 in][]{2LAC}, and the 3LAC  \citep[Fig.\ 10 in][]{3LAC}. The distributions of spectral indices of the entire BL Lac and FSRQ blazar samples follow a pattern, with large scatter, described by the relation $\Gamma_\gamma = d - k \log \nu_{14}$, where $\nu_{s} = 10^{14}\nu_{14}$ Hz. 
For the entire sample of FSRQs and BL Lac objects, the value $k = 0.18\pm0.03$ is found in \citet{3LAC}. A similar function, with different values of $d$ and $k$, apply to the $\Gamma_\gamma$ vs.\ $\nu_{\rm C}$ data. The spectral-index distribution of BL Lac objects with unknown redshift  is generally consistent with the distribution of BL Lac objects with known redshift \citep{2LAC,3LAC}. 

%The coefficients $a_1$ and $b_1$ of the entire Fermi-LAT blazar samples and subsample (e.g., FSRQ vs.\ BL Lac vs.\ SSRQ vs.\ radio galaxies) can be derived from the data for comparison with models.
%Would be useful to provide a fit to Fig. 10 in the form of Gamma_\gamma = a_1 - b_1 log(v_14)

In this paper, we use an equipartition blazar modeling approach \citep{cer13,der14a} {bf
assuming a log-parabole electron energy distribution (EED)} to explain the blazar spectral-index diagrams. In Section 2 we derive analytic Thomson-regime expressions for the relationship between 
$\Gamma_\gamma$ and $\nu_s$, depending on whether the $\gamma$ rays are made through external Compton (EC) or synchrotron self-Compton (SSC) processes. Because of the equipartition relations, the expressions 
depend on $b$, $\nu_{s}$, variability time $t_{var}$, bolometric isotropic synchrotron luminosity $L_{syn}$, an equipartition parameter $\zeta_e$ and a radiation parameter $\zeta_s$. The simpler $\Gamma_\gamma$ vs.\ $\nu_{\rm C}$ expressions are also obtained. The derived analytic relations, { confirmed by numerical modeling,} are shown in Section 3 to be in { general} accord with the blazar spectral-index diagram data, whether external radiation fields in the jet environment are present or absent. { The effects of a log-parabola EED with a low-energy power-law component are also considered.}

In Section 4, application of the equipartition model to the BL Lac object Mrk 501 is demonstrated, and effects of departures from equipartition are evaluated. Trends in spectral-index behavior with other observables constrained can be tested with correlated Fermi-LAT and multiwavelength data, and how this work relates to the blazar sequence and blazar divide, and blazars as UHECR sources, are discussed in Section 5. { The work is summarized in Section 6.}

Appendix A gives a Thomson-regime derivation of the SSC spectrum with a log-parabola electron distribution,
and Appendix B gives a jet power-analysis. { There we show that the 
assumption that blazars are typically observed at the Doppler beaming  angle 
may have led \citet{ghi14} to overestimate the absolute jet power.}
Indeed, out-of-equipartition models are ultimately constrained by  
demands for power.

\section{Equipartition Blazar Modeling with Log-Parabola Electron Energy Distribution}

A standard blazar-jet model, treated in innumerable blazar spectral modeling papers \citep[see][for review]{bhk12}, starts with magnetized plasma that is ejected at relativistic speeds along the poles of a rotating black hole.\footnote{The shock-in-jet model of \citet{mg85} provides an alternate approach that could apply to the $\ll 10^{12}$ Hz radio regime that often remains unfit in the standard model described here.} The jet plasma, which entrains thermal and nonthermal particles in a hypothetical tangled and randomly oriented magnetic field, is a source of escaping photons, and potentially also of escaping cosmic rays and neutrinos. The jet power is extracted from the mass energy of accreting matter and/or the rotational energy of the black hole itself. The collimated relativistic plasma outflow, an exhaust byproduct of the energy generated by the black-hole engine, is usually attributed to
processes taking place in the magnetosphere of the rotating black hole.
The  polarized broad-band synchrotron radiation emitted by an energetic EED
(which could also contain positrons) is boosted by the Doppler effect along the jet axis, so that rapidly variable
jet synchrotron radiation can be detected by Earth-based observatories from large redshift ($z\gg 1$) sources.

The jet electrons also Compton scatter ambient photons to $\gamma$-ray energies. Besides the accompanying SSC emission from target synchrotron photons \citep[e.g.,][]{mgc92,bm96},  EC $\gamma$ rays are made when the nonthermal jet electrons scatter photons from external radiation fields. Depending on jet Doppler factor $\delta_{\rm D}$ and BLR cloud parameters, the direct accretion-disk radiation field dominates the external radiation field of a powerful FSRQ at $\ll 10^3$ Schwarzschild radii, BLR fields are strongest within $\sim 0.3$ pc \citep{dsm92,sbr94,ds02}, while at the pc scale and beyond, infrared radiation from a surrounding IR-emitting dust torus would have the largest  energy density of all ambient radiation fields \citep{bla00,sik09,gt09} in the inner blazar-jet environment.
%Except in unusual situations, a shell geometry and a comoving spherical blob geometry yield similar predictions.

The form of the nonthermal EED
is often treated by either assuming a nonthermal
injection spectrum of leptons that evolves in response to adiabatic
and radiative losses, or by assuming a form for the average steady-state EED
in the radiating jet plasma. Adopting the latter approach, we assume
that the 3 parameter log-parabola function
\begin{equation}
\g^{\prime 2} N_e^\prime(\gp)=[\gamma_{pk}^{\prime 2}N^\prime_e(\g^\prime_{pk})]({\g^\prime\over\g^\prime_{pk}})^{-b\log({\g^\prime\over \gamma_{pk}^\prime})}
\equiv   K^\prime y^{- b\log y}\;
\label{g2Ng}
\end{equation}
provides an approximate description of the nonthermal lepton spectrum. Here $y\equiv \gamma^\prime/\gamma^\prime_{pk}$,
$\gamma_{pk}^\prime$ is the peak, or principal, Lorentz factor of the fluid-frame EED, eq.\ (\ref{g2Ng}).
The value of $K^\prime$ can be related to either the total particle number or total comoving particle energy \citep{der14a}; in the latter
case, $K^\prime = {\cal E}^\prime_e/m_ec^2 \sqrt{\pi\ln 10/b}$, where ${\cal E}^\prime_e$ is the nonthermal electron energy of the blob.
%, with $I_1(b) = \sqrt{\pi\ln 10/b}$

The continuously curving EED given by a log-parabola function derives from stochastic acceleration
processes with radiation and escape \citep[see, e.g.,][]{mas04,bld06,tra07,tra11,sp08}.
 With this form of the EED, GeV breaks in FSRQs and blazars with $\nu_{pk}^{syn} \lesssim 10^{14}$ Hz are shown to arise from the onset of Klein-Nishina effects when scattering BLR photons \citep{cer13,ack10,tg08}, and to give \citep{der14a} reasonable fits to four epochs of quasi-simultaneous multiwavelength observations of 3C 279 \citep{hay12}. As we show below, this approach also gives good fits to the SED of Mrk 501,
though the best fits are achieved with an electron distribution out of equipartition with the magnetic field.
%This spectral modeling approach is illustrated in the next section by application to Mrk 501.

The comoving synchrotron luminosity
\begin{equation}
L_{syn}^\prime = c \sigma_{\rm T} {B^{\prime 2}\over 6\pi}  \int_1^\infty d\gamma^\prime \gamma^{\prime 2} N_{e}( \gamma^\prime)\;
\label{Lsyn}
\end{equation}
implies, using Eq.\ (\ref{g2Ng}) and a $\delta$-function approximation for the synchrotron photon with average dimensionless energy $\e_{syn} = (3/2)\delta_{\rm D} (B^\prime/B_{cr})\gamma^{\prime 2}$ \citep{dm09}, the received synchrotron luminosity spectrum
\begin{equation}
\e L_{syn}(\e ) = \upsilon x^{1-\hat b \ln x}= \upsilon ({\epsilon\over\e_{pk}})^{{1\over 2}-{b\over 4}\log ({\e/\e_{pk}})}\;,
\label{Lsyne}
\end{equation}
where $x = \sqrt{\e/\e_{pk}}$, $\upsilon = f_3 L_{syn}$, and $f_3^{-1} = 2\cdot 10^{1/4b}\sqrt{\pi \ln 10/b}$ \citep{der14a}.
Thus the effective log-parabola width parameter $b_{sy}$ for the synchrotron spectrum is given by $b_{sy} = b/4$ in the $\delta$-function
approximation, and $b_{sy} \cong b/5$ when using the full Thomson cross section \citep{mas06,pag09}.
The peak synchrotron frequency $\epsilon_{pk} = (3/2)\delta_{\rm D}(B^\prime/B_{cr})\gamma_{pk}^{\prime 2}$.  The slope of the $\e L_{syn}(\e )$ spectrum is
\begin{equation}
\alpha_\nu \equiv {d \ln[\e L_{syn}(\e )]\over d\ln \e } = {1\over 2} [1- b \log (\e/\e_{pk})]\;.
\label{alphanu}
\end{equation}
\citep{mas04a}. Because the nonthermal electron energy-loss rate from synchrotron processes scales quadratically  with electron Lorentz factor $\gamma^\prime$, the synchrotron spectrum from a log-parabola distribution of electrons has a $\nu L_\nu$ synchrotron peak energy $\epsilon_s = h\nu_s/m_ec^2$ that is shifted to higher values than $\epsilon_{pk}$. Eq.\ (\ref{alphanu}) shows that $\epsilon_s = 10^{1/b} \e_{pk}$   \citep{mas06}.
 
\begin{table}[t]
\begin{center}
\caption{Dependences of $\delta_{\rm D}$, $B^\prime$, and $\gamma_{]pk}^\prime$$^a$}
\begin{tabular}{cccccccccccc}
\hline
${\rm }$ & Coef. & $L_{48}$ & $\nu_{14}$ &  $t_4$ & & $\zeta_s$  & $\zeta_e$ & & $f_0$  & $f_1$ & $f_2$\\
$$ &  & $ $   &   &   &   &    & & &  &  \\
\hline
\hline
 &  &  &   &   &   &  &  & & &  & \\
$\delta_{\rm D}$ & 17.5 & $~~3/16$ & $~~1/8$ & $-1/8$ & & $-7/16$ & $~~1/4$ & & $-7/16$ & $-1/4$ & $-1/8$ \\
% &  &  &  &  & &   & &   \\
$B^\prime({\rm G})$ & 5.0 &  $-1/16$  &  $-3/8$ & $-5/8$ &  & $~13/16$ & $-3/4$ & & $~13/16$ &  $~~3/4$ & $~~3/8$ \\
$\gamma_{pk}^\prime$ & 523  &  $-1/16$ & $~~5/8$ & $~~3/8$ & & $-3/16$ & $~~1/4$ & & $-3/16$ & $-1/4$ & $-5/8$ \\
 &  &  &  &  & &   & &   \\
${\cal E}^b$ & $1.4$ &  $5/16$  &  $-1/8$ & $1/8$ &  & $-1/16$  & $-1/4$ & & $-11/16$ &  $~~1/4$ & $~~1/8$ \\
$L_{jet,B}^c$ & $4$ &  $5/8$  &  $-1/4$ & $1/4$ &  & $-1/8$ & $-1/2$ & & $-1/8$ &  $~~1/2$ & $~~1/4$ \\
\\
\hline\end{tabular}
\label{table1}
\end{center}
$^a$ So, e.g., $\delta_{\rm D} \cong 17.5 L_{48}^{3/16}(\nu_{14}/f_2t_4)^{1/8} (f_0\zeta_s)^{-7/16}(\zeta_e/f_1)^{1/4}$, etc. \\
$^b$ ${\cal E} =E_{max}(10^{20}~eV)/Z$\\
$^c$ Absolute power in magnetic field, units of $10^{44}~$erg s$^{-1}$\\
%\label{table1}
\end{table}

Table \ref{table1} shows the various dependencies of blob properties on the observables $L_{48}$, 
$\epsilon_s$ (or $\nu_{14}$), $t_{var}$ and $b$, and on the
equipartition factor $\zeta_e$ and radiative factor $\zeta_s$. The factor $\zeta_e$ is the ratio of nonthermal electron energy density $u^\prime_e$ to magnetic-field energy density $u^\prime_{B^\prime}= B^{\prime 2}/8\pi$, 
and $\zeta_s$ is the ratio the jet-frame synchrotron photon energy density and $u^\prime_{B^\prime}$.  
In the blob scenario, the geometry factor $f_0 = 1/3$. The $b$-dependent factors are 
$f_1 = 10^{-1/4b}$, $f_2 = 10^{1/b}$, and $f_3 = (2\cdot 10^{1/4b} \sqrt{\pi \ln 10/b})^{-1}$ \citep{der14a}, so $\epsilon_s = f_2 \e_{pk}$.

The similarity of the underlying physics of the synchrotron and Compton processes \citep{bg70} means that an expression like eq.\ (\ref{alphanu}) holds for Compton scattering in the Thomson regime, except now $\epsilon_{pk}$ is replaced by a corresponding peak photon energy for EC and SSC processes \citep[e.g.,][]{pag09}.  In the EC case,  $\epsilon_{pk, EC} = (4/3)\delta_{\rm D}^2 \epsilon_0 \gamma_{pk}^{\prime 2}$, assuming an isotropic monochomatic external radiation field with energy $\e_0$ and energy density $u_0$.  From the equipartition relations \citep{der14a} shown in Table 1 for the photon spectral index $\Gamma_\gamma = 2-\alpha_\nu$, we find that the photon index for EC processes is given by
\begin{equation}
\Gamma_{\gamma}^{EC} \cong {17\over 8} + {b\over 2}\,\log\left({f_0^{5/4}E_{GeV} \zeta_s^{5/4}}\over \epsilon_{Ly\alpha}t_4^{1/2} \zeta_e L_{48}^{1/4} \right )
- {3b\over 4}\log \nu_{14}\;. \\
%= {3\over 2} + {b\over 2}\,\log\left({E_{GeV}{f_{EC}} \zeta_{EC}\over \epsilon_{Ly\alpha}t_4^{1/2} L_{48}^{1/4}}\right)
%- {3b\over 4}\log \nu_{14}\;.
\label{GammagammaEC}
\end{equation}
Here $E_{GeV}$ is the effective detection energy in GeV, and $\e_0 = 2\times 10^{-5}\epsilon_{Ly\alpha }$ for Ly $\alpha$/BLR scattering.
%$f_{EC} = f_0^{5/4} f_1 f_2^{3/2}$,
%and $\zeta_{EC} = \zeta_s^{5/4}/\zeta_e $. 
A nominal value of $E_{GeV} = 1$ is chosen because the Fermi-LAT is most sensitive at $\approx 1$ GeV \citep[Fig.\ 18 in][for a $\Gamma_\gamma= 2.2$ source spectrum]{1FGL}. The dependence of $\Gamma_\gamma$ on $E_{GeV}$ can be studied by analyzing Fermi-LAT data in discrete energy ranges.  

Scattering the dusty torus emission, with IR photon energies corresponding to $\epsilon_{Ly\alpha }\sim 0.02$, implies a Thomson spectrum softer by $\Delta\Gamma_\gamma \cong 0.85 b$, because $\gamma$-ray photons at a given observing energy are produced in the softer part of the Compton-scattered spectrum when the target photons have lower energies.       If equipartition is instead made to total particle energy density $u^\prime_{tot}$ according to the factor $\zeta_{eq} =u^\prime_{tot}/u^\prime_{B^\prime}$, then $\zeta_e = \zeta_{eq}/(1+\eta_{bl})$ and $\eta_{bl} = u^\prime_{baryons}/u^\prime_{e}$ is the baryon loading, and $u^\prime_{baryons}$ is the internal energy density in protons and ions   \citep[][and Appendix B]{dmi14}.

The specific spectral synchrotron luminosity, 
from eq.\ (\ref{Lsyne}) in the $\delta$-function approximation and results of \citet{der14a}, 
is given by 
\begin{equation}
\e L_{syn}(\e,\Omega) = f_3 N_e \, {4\over 3}\, c \sigma_{\rm T} {B^{\prime 2}\over 8\pi }\,\gamma_{pk}^{\prime 2} \delta_{\rm D}^4
x^{1-b\log x}\;,
\label{eLsynesOmegas}
\end{equation}
where
$x = \sqrt{\e/\e_{pk}}$, and $\e_{pk} = 4 \delta_{\rm D} B^\prime \gamma_{pk}^{\prime 2}/3B_{cr}$.
The specific spectral $\gamma$-ray luminosity in the Thomson regime for 
a jet traveling through an external isotropic,
monochromatic radiation field with frequency $m_e^2 \epsilon_0/h$ and energy density $u_0$, in units of $m_ec^2$ cm$^{-3}$, using a $\delta$-function approximation for Thomson scattering, is
%the isotropic Thomson expression for the Compton cross section \citep[eqs.\ (6.71) and (6.112) in][]{dm09}, is 
\begin{equation}
\e L_{EC}(\e,\Omega) \cong \,f_3 N_e \, {4\over 3}\, c \sigma_{\rm T} u_0\,\gamma_{pk}^{\prime 2} \delta_{\rm D}^6
{\tt v}^{1-b\log {\tt v}}\;,
\label{eLECesOmegas}
\end{equation}
where ${\tt v} \equiv \sqrt{\e/\e_{pk,EC}}$ and $\e_{pk,EC} = (4/3)\delta_{\rm D}^2 \gamma_{pk}^{\prime 2}\e_0$.
The technique of \citet{gkm01} is used to derive this expression. The ratio of the spectral synchrotron
and Thomson luminosities at their respective
peak frequencies is $\delta_{\rm D}^2 u_0/u^\prime_{B^\prime }$.

For the SSC process, the combined effects of the widths of both the EED and the target synchrotron photon spectrum will broaden the Compton-scattered photon spectrum such that its effective width in the Thomson regime is obtained by replacing $b$ by $b_{SSC} = b/2$ in eq.\ (\ref{alphanu})  \citep{pag09} and replacing $\e_{pk}$ by $\e_{pk,SSC} = 2\delta_{\rm D} (B^\prime/B_{cr}) \gamma_{pk}^{\prime 4}$, giving
$$\Gamma_{\gamma}^{SSC} =  {65\over 32} + {b\over 4}\,\log\left( {6.5\times 10^3 E_{GeV}f_{0}^{3/8} \zeta_{s}^{3/8}L_{48}^{1/8}\over \zeta_e^{1/2} t_4^{3/4} }\right)$$
\begin{equation}
%+ {b\over 4}\,\log\left( {6.5\times 10^3 E_{GeV}{f_{SSC}} \zeta_{SSC}L_{48}^{1/8}\over t_4^{3/4} }\right)
- {9b\over 16}\log \nu_{14}\;,
\label{GammagammaSSC}
\end{equation}
from Table 1.
%Here
%$f_{SSC} = f_0^{3/8} f_1^{1/2} f_2^{9/4}$, and
%$\zeta_{SSC} = \zeta_s^{3/8}/\zeta_e^{1/2}$.
Note that the $\nu L_\nu$ peak SSC frequency is a factor $10^{2/b}$ larger than $\e_{pk,SSC}$.
The SSC expression is justified by a more detailed derivation in Appendix A.
%and that some minor simplifications can be made
%by expanding $\zeta_{EC}$ and $\zeta_{SSC}$in eqs.\ (\ref{GammagammaEC}) and (\ref{GammagammaSSC}),respectively.
The uncertainty $\Delta \Gamma_\gamma$ in the spectral index related to geometrical
uncertainties can be estimated by letting $f_0$ range from unity
for a blast-wave shell geometry to $f_0 = 1/3$ for a comoving spherical-blob geometry.
From eqs.\ (\ref{GammagammaEC}) and (\ref{GammagammaSSC}), one can see that this translates into
an uncertainty $\Delta \Gamma_\gamma^{EC} \cong 0.30 b$ for EC processes and an
 uncertainty $\Delta \Gamma_\gamma^{SSC} \cong 0.04 b$ for SSC processes.

We also derive the Thomson-regime expressions
\begin{equation}
{\Gamma^{EC,\gamma}_\gamma = {2} +{b\over 2}\log (2.4 E_{GeV}) - {b\over 2}\log \nu_{23}\;}
\label{Gg}
\end{equation}
for $\Gamma_\gamma$ vs.\ $\nu_C$ in EC processes, and
\begin{equation}
{\Gamma^{SSC,\gamma}_\gamma = 2 +{b\over 4}\log (2.4 E_{GeV}) - {b\over 4}\log \nu_{23}\;}
\label{GgSSC}
\end{equation}
for $\Gamma_\gamma$ vs.\ $\nu_C$ in SSC processes. Here $\nu_{23} = \nu_C/10^{23}$ Hz is the
 peak frequency of the Compton component of the $\nu L_\nu$ SED. Note that eq.\ (\ref{Gg})
is independent of the target photon energy, because the expression assumes that the EED
and Doppler factor are adjusted to produce a Compton-scattered $\gamma$-ray spectrum
that peaks at $\nu_{\rm C}$.

\begin{figure}[lt]
\begin{center}
 \includegraphics[width=3.5in]{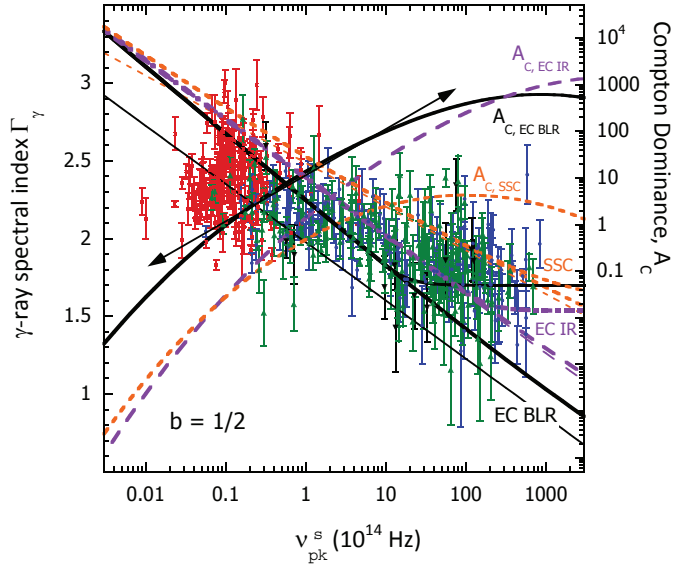} %Two column width = 3.3in
 \caption{\small
Data are the $> 100$ MeV photon spectral index values $\Gamma_\gamma$ as a function of peak synchrotron frequency $\nu_s$
for blazars from the 2LAC \citep{2LAC}.
Red, blue, green, and black symbols identify, respectively, FSRQs, BL Lac objects with redshifts, BL Lac objects without redshifts, and blazars with data too poor to determine if the source is an FSRQ or a BL Lac object. 
%Log-parabola equipartition-model predictions using standard parameter values, eq.\ (\ref{stanpar}) with $b = 1/2$, are shown for comparison with the data. 
{\it Left:} Curves labeled by EC BLR, EC IR and SSC for EC processes with BLR photons, EC processes with IR photons and SSC processes, respectively,  show  $\Gamma_\gamma$ vs.\ $\nu_s$ predictions of the log-parabola equipartition model using standard parameters given by eq.\ (\ref{stanpar}).
Also, $\epsilon_0 = 2\times 10^{-5}$ and $u_0= 10^{-2}$ erg cm$^{-3}$ for Ly$\alpha$, and $\epsilon_0 = 4.6\times 10^{-7}$ and $u_0= 10^{-3}$ erg cm$^{-3}$ for the $\sim 1000$ K IR radiation.
Thick curves give numerical calculations, and thin curves show analytic results, from eqs.\ (\ref{GammagammaEC}) and (\ref{GammagammaSSC}). { The thick curves that approach constant values at large $\nu_s$ are numerical predictions for the power-law, log-parabola model, eq.\ (\ref{PLLP}). }
%with BLR photons (EC BLR), EC with infrared radiation (EC IR), and SSC processes, as labeled. 
%The light curves with the same line styles are values derived from the analytic expressions, eqs.\ (\ref{GammagammaEC}) and (\ref{GammagammaSSC}). 
{\it Right:} Compton-dominance ${\cal A}_{\rm C}$ as a function of $\nu_s$ for EC BLR, EC IR, and SSC processes, as labeled. 
%giving approximately the ratio of the bolometric $\gamma$-ray and synchrotron luminosities,
% reach extreme values for EC IR and EC BLR processes when $\nu_s\gg 10^{15}$ Hz.  
The line with arrows has a slope of $+1$ in the ${\cal A}_{\rm C}$ vs.\ $\nu_s$ plane. 
}
\label{fig1}
\end{center}
\end{figure}

\section{Modeling the Blazar Spectral-Index Diagram}

Fig.\ \ref{fig1} shows measured values of Fermi-LAT spectral index $\Gamma_\gamma$ from the 2LAC \citep{2LAC} derived from a single power-law fit to the complete data set in the 0.1 -- 100 GeV range for sources with $TS>25$.\footnote{Energy flux is derived in 5 energy bands in intervals defined by 0.1, 0.3, 1, 3, 10 and 100 GeV.} The red, blue, green, and black data symbols correspond, respectively, to $\gamma$-ray sources detected with the Fermi-LAT that have been associated with FSRQs, BL Lac objects with and without redshifts, and blazars with optical data too poor to determine if the source is an FSRQ or BL Lac.  

From  inspection of the plot, it is clear that a function of the form $\Gamma_\gamma = d - k \log \nu_{14}$ will provide a reasonable description of the data. For the entire FSRQ and BL Lac sample, but excluding other blazar candidates, values of $k = 0.18\pm 0.03$ and $d =2.25\pm 0.04$ are deduced in the 3LAC \citep{3LAC}.  Comparing this value with the analytic expressions, eqs.\ (\ref{GammagammaEC}) and (\ref{GammagammaSSC}), a larger value of $b$ is implied for SSC processes compared to EC processes, but in both cases consistent with $b \approx 1/3$. 
%However, spectral modeling shows that 

The typical value of $b$ can also be deduced from the average nonthermal blazar synchrotron SED, when fit with an expression of the form of Eq.\ (\ref{Lsyne}).  
%Visual inspection of the Bright Fermi Blazar SED paper \citep{SED} indicates that $0.5 \lesssim b_{sy} \lesssim 1$, implying $b \approx b_{sy}/5 \approx 0.1$ -- 0.2. 
From X-ray analysis of Beppo-SAX data on Mrk 501, \citet{mas04} finds values of $b_{sy}$ ranging from 0.12 -- 0.33, implying
 a  corresponding log-parabola width parameter $b\gtrsim 0.5$. Narrow bandwidth modeling of X-ray synchrotron emission from Mrk 421 gives $b_{sy} \cong 0.3$ -- 0.5 \citep{tra07}, though 
values of $b_{sy} = 0.17 \pm 0.02$ (2006 15 July pointing), 
$b_{sy} = 0.11 \pm 0.02$ (2006 April 22 pointing), and 
$b_{sy} = 0.08 \pm 0.03$ (2006 June 23 pointing) are obtained in more complete
joint XRT-BAT analysis \citep{tra09},  consistent with an electron distribution with $b \cong 5 b_{sy} \approx 0.5$.  
\citet{chen14} finds that  $b_{sy}$ is distributed in the range $0.05 \lesssim b_{sy} \lesssim 0.25$, implying $0.25 \lesssim b \lesssim 1.25$.
More importantly, he finds a dependence of $b_{sy}$ on $\nu_s$, which we discuss further in Section \ref{sec:ds}. 
%%%Need to see if this is the first presentation of this result
The values of $b$ deduced from spectral modeling tend to be larger than obtained from the slope implied by the spectral-index diagram. 

\subsection{ Standard parameters in log-parabola model}

To compare the log-parabola equipartition model with data, we adopt a {standard} parameter set, and take
\begin{equation}
b=1/2,\;t_4 = L_{48} = \zeta_{e} = \zeta_{s} = E_{GEV} = 1\;.
\label{stanpar}
\end{equation}
The reasoning driving the choice of
the standard variability time scale is that
 the masses of supermassive black holes powering blazars---both FSRQs and BL Lacs---are typically of the order $\sim 10^9 M_\odot$.
The value  $t_4 \cong 1$ or $t_{var} \cong 3$ hr corresponds to the light-crossing time across a size equal to the Schwarzschild radius of a $\sim 10^9 M_\odot$ black hole, though of course shorter variability time scales  have been
recorded during spectacular outbursts of BL Lac objects, including Mrk 421 \citep{fos08MRK421}, Mrk 501 \citep{alb07Mrk501}, and PKS 2155-304 \citep{aha07PKS2155},  not to mention the extraordinary VHE outburst observed with the MAGIC telescope from the FSRQ PKS 1222$+$216 with $t_{var} \sim 10$ m \citep{PKS1222}.
The isotropic synchrotron luminosity $L_{syn}$ can exceed the Eddington limit $L_{\rm Edd}$, though $L_{\rm Edd}$ is presumably the upper limit to the persistent absolute jet power (see App.\ B). Standard values $L_{48}\sim 0.1$ -- 1 and $L_{48}\sim 10^{-2}$ -- $10^{-3}$ are typical of powerful FSRQs and BL Lac objects, respectively.
At the other side of the time domain, $t_{var} \sim 10^5$ -- $10^6$ s may be compatible with quiet times of blazars.

Fig.\ 1 shows analytic results of  Eqs.\ (\ref{GammagammaEC}) and (\ref{GammagammaSSC}) for $\Gamma_\gamma$ as a function of $\nu_{\rm pk}^s =  \nu_s$, using the standard parameter set. Results of numerical calculations, obtained by modifying the code used in \citet{der14a}, are also shown. The dimensionless photon energies for the BLR and IR photons used in the model are $\epsilon_0 = 2\times 10^{-5}$ (i.e., 10.2 eV) for BLR photons and $\epsilon_0 = 4.6\times 10^{-7}$ for warm IR torus dust emission described by an $\approx 1000$ K greybody spectrum with $\approx 15$\% covering factor, giving an energy 
density of $\approx 10^{-3}$ erg cm$^{-3}$. The analytic results are shown by the thin lines. The numerical results are shown by the thick curves.
% with extrapolations to smaller and larger values of $\nu_{14}$ shown by dots, and explained below.
As can be seen, the analytic SSC and EC IR results are in reasonable agreement with the numerical calculations, whereas the analytic EC BLR results do not agree with the numerical results. Klein-Nishina effects already make themselves felt strongly for target BLR photons scattered to 1 GeV, but only weakly for target IR photons scattered to 1 GeV, as is clear by noting that KN effects set in at photon energies $E_\gamma\gtrsim m_ec^2/12\epsilon_0 \approx 100$ GeV for 1000 K photons, and $E_\gamma \approx 2$ GeV for Ly $\alpha$ photons. The Thomson-regime expressions are harder than the numerical curves because of the Klein-Nishina softening.

{  Fig.\ 1 also shows the effects of a low-energy power-law extension of the EED on the spectral-index diagrams.
In such a power-law log-parabola (PLLP) model with a low-energy cutoff Lorentz factor $\gamma^\prime_{min}$ \citep{yan13,pyz14}, the EED distribution extends eq.\ (\ref{g2Ng}) by two parameters to take the form
\begin{equation}
\gamma^{\prime 2}N_e^\prime(\gamma^\prime) = K_e^\prime \,[y^{2-s}H(y;y_{\ell},1)+y^{2-s-r\log y} H(y-1)]\;.
\label{PLLP}
\end{equation}
Here $s$ is the power-law spectral index of the low-energy component,  $r$ is a log-parabola width parameter, 
and $y_{\ell} = \gamma^\prime_{min}/\gamma_{pk}$. The Heaviside functions are defined such that $H(u) = 1$ when $u\geq 0$ and $H(u) = 0$ otherwise, and $H(u;a,b) = H(u-a)H(b-u)$.
The theoretical basis for the form of eq.\ (\ref{PLLP}) is discussed below.  Results are shown for $s = 2$ and $y_\ell \ll 1$, 
in which case $r\rightarrow b$, reducing the PLLP model to a 3-parameter model. 
}

\subsection{ Compton dominance}

The numerical results for this particular set of parameters are seen to follow the trend of much of the data.
Virtually no FSRQs are observed, however, with  $\nu_{14} > 1$. To obtain some insight into this, we calculate the Compton dominance ${\cal A}_{\rm C}$ for our model, defined here as the ratio of the 100 MeV -- 100 GeV $\gamma$-ray luminosity to the bolometric synchrotron luminosity. It is calculated from the relation
\begin{equation}
{\cal A}_{\rm C} \equiv {L_\gamma(100~{\rm MeV})\over \alpha_\nu L_{syn}}\;[({100\over E_{GeV}})^{\alpha_\nu} - ({0.1\over E_{GeV}})^{\alpha_\nu}]
\label{AC}
\end{equation}
where $\alpha_\nu$ and $L_\gamma(100~{\rm MeV})$ are, respectively, the $\nu L_\nu$ spectral index and luminosity calculated at $E_{GeV}$ GeV.
Note that a more detailed and time-intensive calculation would integrate the blazar SED to determine ${\cal A}_{\rm C}$.
%Because we let $\zeta_s = 1$, the{}

The Compton dominance depends on the energy density of the surrounding radiation fields. For definiteness, we have taken $u_{BLR} = 10^{-2}$ erg cm$^{-3}$ and $u_{IR} = 10^{-3}$ erg cm$^{-3}$ in our calculations. Note that ${\cal A}_{\rm C}$ scales approximately linearly with $u_0$.
As ${\cal A}_{\rm C}$ becomes progressively smaller, the corresponding blazars becomes progressively less detectable as $\gamma$-ray sources. So solutions should be restricted to a minimum value of ${\cal A}_{\rm C}$. Solutions should also be restricted at large values of  ${\cal A}_{\rm C}$, because Compton drag on the jet becomes a strongly limiting factor, as discussed more in Section \ref{sec:ds}. 
%If excessive Compton drag restricts
%HSP objects to regions of low-external radiation density, 
Regions where $0.1 \lesssim {\cal A}_{\rm C} \lesssim 30$ 
may favor LSP blazars to be FSRQs, ISP blazars to 
the EC  BLR, EC IR, and SSC solutions in Fig.\ \ref{fig1}, as these
values bracket measured values of the Compton dominance \citep[Fig.\ 7 in][]{fin13}.
%The upper limit reflects not only the largest values of ${\cal A}_{\rm C}$ seen in blazar physics [CHECK], but may violate the underlying assumptions of the equipartition model, as discussed further below.
%The equipartition solution always produces

\begin{figure}[t]
\begin{center}
 \includegraphics[width=3.5in]{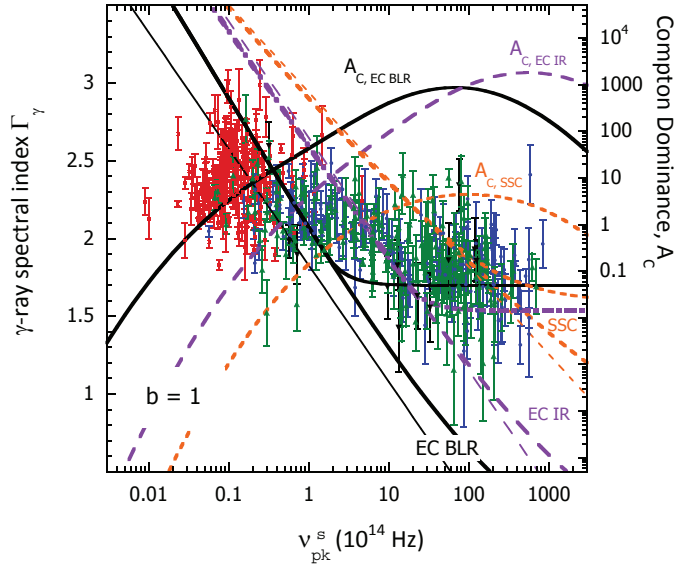}
 \caption{Same as Fig.\ \ref{fig1}, except that $b = 1$. Heavy and light downward-going curves are the numerical and analytic equipartition model predictions, respectively, and upward going curves show  Compton dominance for EC BLR, EC IR and SSC processes { for the log-parabola EED, eq.\ (\ref{g2Ng}).} {  The thick curves approaching constant values at large values of $\nu_s$ correspond to spectral-index predictions of the PLLP model, eq.\ (\ref{PLLP}), with a $-2$ number index of the low-energy EED.} 
%The line with arrows has a slope of $+1$ in the ${\cal A}_{\rm C}$ vs.\ $\nu_s$ plane. }
%[Still have to put in dust curves and improve here and in Figs. \ref{fig3} and \ref{fig4}. Note that line styles are not consistent.] 
}
\label{fig2}
\end{center}
\end{figure}

\begin{figure}[t]
\begin{center}
 \includegraphics[width=3.5in]{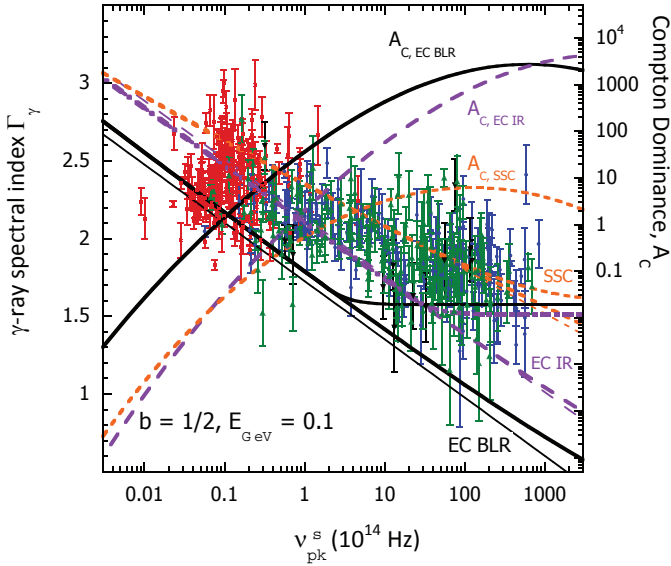}
 \caption{Same as Figs.\ \ref{fig1} and \ref{fig2}, except that $E_{GeV} = 0.1$.  }
\label{fig3}
\end{center}
\end{figure}

\begin{figure}[t]
\begin{center}
 \includegraphics[width=3.5in]{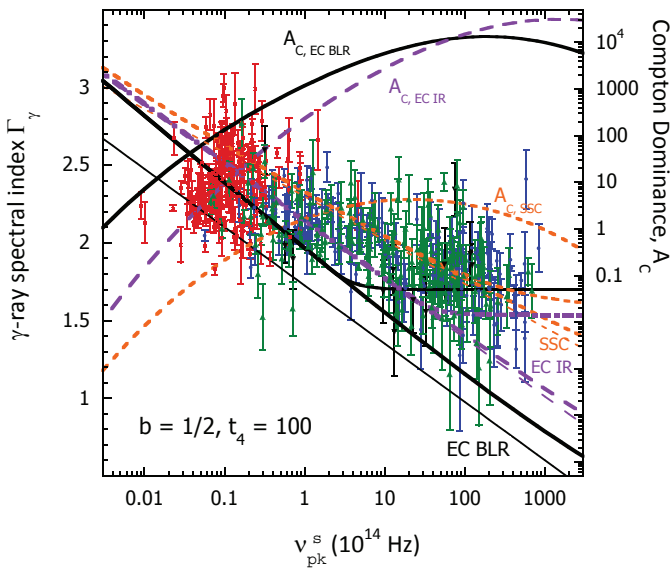}
 \caption{Same as Figs.\ \ref{fig1} and \ref{fig2}, except that $t_{4} = 100$. }
\label{fig4}
\end{center}
\end{figure}

In Fig.\ \ref{fig1}, we calculate three models in the $\Gamma_\gamma$ vs.\ $\nu_s$ plane corresponding to complete
dominance either of Ly $\alpha$ BLR radiation (EC BLR), IR radiation from the dusty torus (EC IR), or internal synchrotron radiation (SSC) as the target photon source.
Restricting the Compton dominance to $0.1\lesssim {\cal A}_{\rm C} \lesssim 30$ suggests that most blazars with $\nu_{14} < 0.1$ have $\gamma$ rays that result from scattered BLR radiation,
while blazars with $0.1 \lesssim \nu_{14}\lesssim 1$ would have a mix of blazars with $\gamma$ rays made by Compton scattering of either BLR or IR photons, or both.
At higher peak synchrotron frequencies, SSC-dominated sources would be most plentiful.

{ The use of a different model, the PLLP EED, eq.\ (\ref{PLLP}) with $s=2$ and $y_\ell \ll 1$, 
is displayed in  Fig.\ \ref{fig1} and subsequent figures by the numerically calculated spectral index curves
that approach constant values of spectral index at $\nu_{14}\gg 1$.  Klein-Nishina effects, described more below, soften
the spectral index below the Thomson regime value of $\Gamma_\gamma = 1.5$. It is interesting that essentially all data
are softer than $\Gamma_\gamma = 1.5$, and that smaller values of external radiation energy density could yield typical
measured Compton dominance values for ISP and HSP blazars with an external Compton $\gamma$-ray component making a 
significant contribution to the SED.}

Figs.\ \ref{fig2} -- \ref{fig4} show how changes in the model parameters affect results. Fig.\ \ref{fig2} shows that a value of $b = 1$ is incompatible with the combined trend of the  data, though a values of $b\cong 1$ may be consistent with sub-populations, e.g., FSRQs. Returning to $b = 1/2$, Fig.\ \ref{fig3} shows the effects of calculating the spectral index at $E_{GeV} = 0.1$, that is, at 100 MeV rather than 1 GeV. Because
the Compton-scattered $\gamma$-ray SED becomes progressively softer at larger $\gamma$-ray energies, the model results in Fig.\ \ref{fig3} are uniformly harder than in Fig.\ \ref{fig1}. The discrepancy between the analytic and numerical results decreases when scattering Ly $\alpha$ radiation because the Klein-Nishina effects on the Compton
cross section are not so great when scattering to 100 MeV as compared to 1 GeV.  The  dependence on detector energy $E_{GeV}$ should clearly show up in Fermi-LAT spectral index diagrams calculated in discrete energy ranges, e.g., 0.3 -- 3 GeV and 3 -- 30 GeV, and should, in a statistical study, discriminate between EC and SSC processes, though correlations between $b$ and $\nu_s$ can hide the effect.

Fig.\ \ref{fig4} shows how a slower variability time, with $t_4 = 100$, affects the equipartition spectral-index
diagram. Compared to the results in Fig.\ \ref{fig1}, the effect of longer
variability times is to harden the spectrum. From eqs.\ (\ref{GammagammaEC}) and (\ref{GammagammaSSC}), the hardening for a factor of 10 longer variability time is $\Delta\Gamma_{\gamma}^{EC} = -b/4$ for EC processes
and $\Delta \Gamma_{\gamma}^{SSC} = - 3b/16$ for SSC processes. Because of the difficulty in measuring $t_{var}$,  the variability effect on spectral index may be too subtle to discriminate between EC and SSC processes. In a statistical sample, however, more rapidly variable sources at equipartition would in general be softer, assuming that there are no underlying correlations between $b$ and $t_{var}$, and that equipartition holds in the various states.

\begin{figure}[t]
\begin{center}
 \includegraphics[width=3.3in]{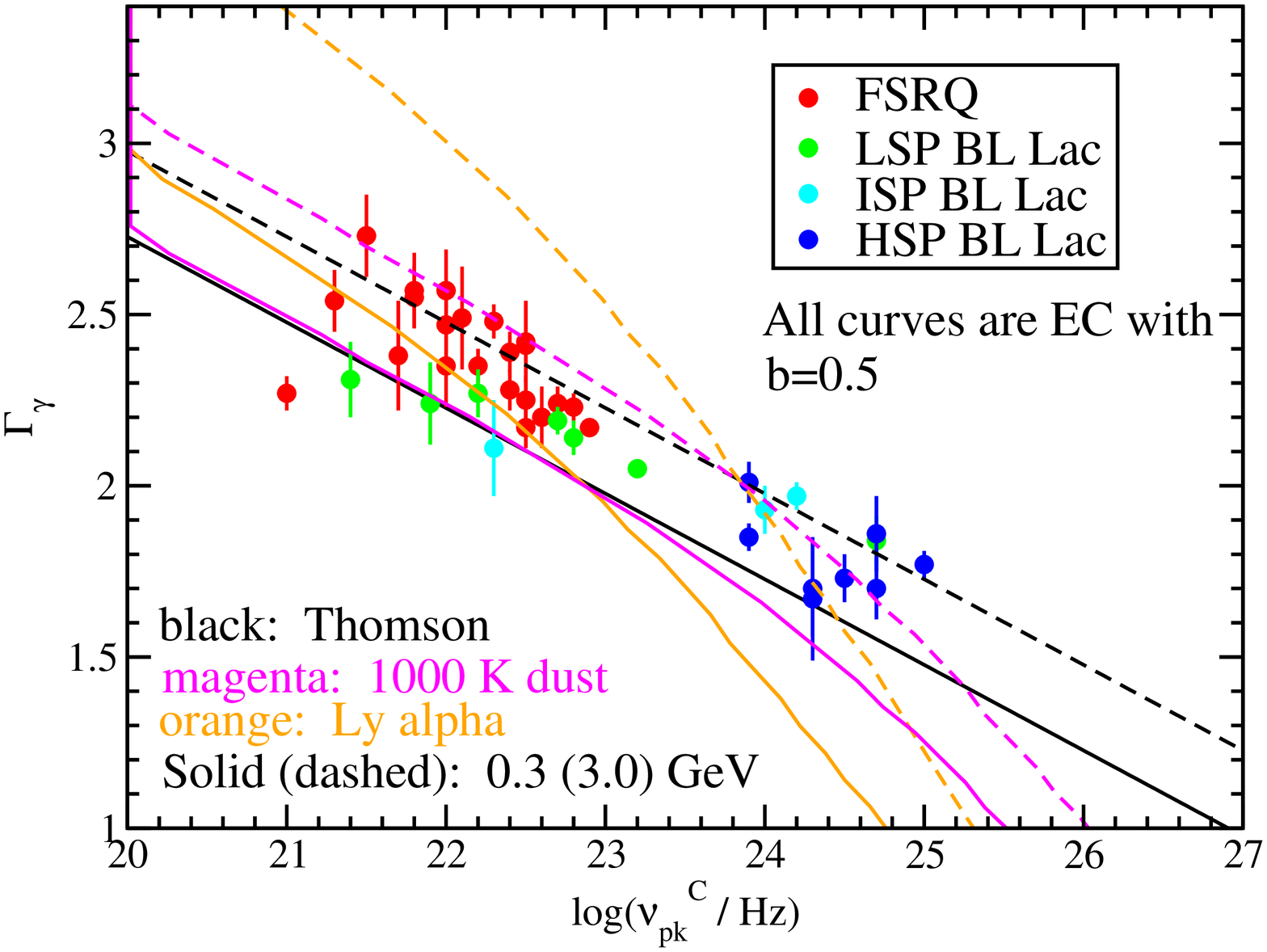}
\vskip0.4in
 \includegraphics[width=3.3in]{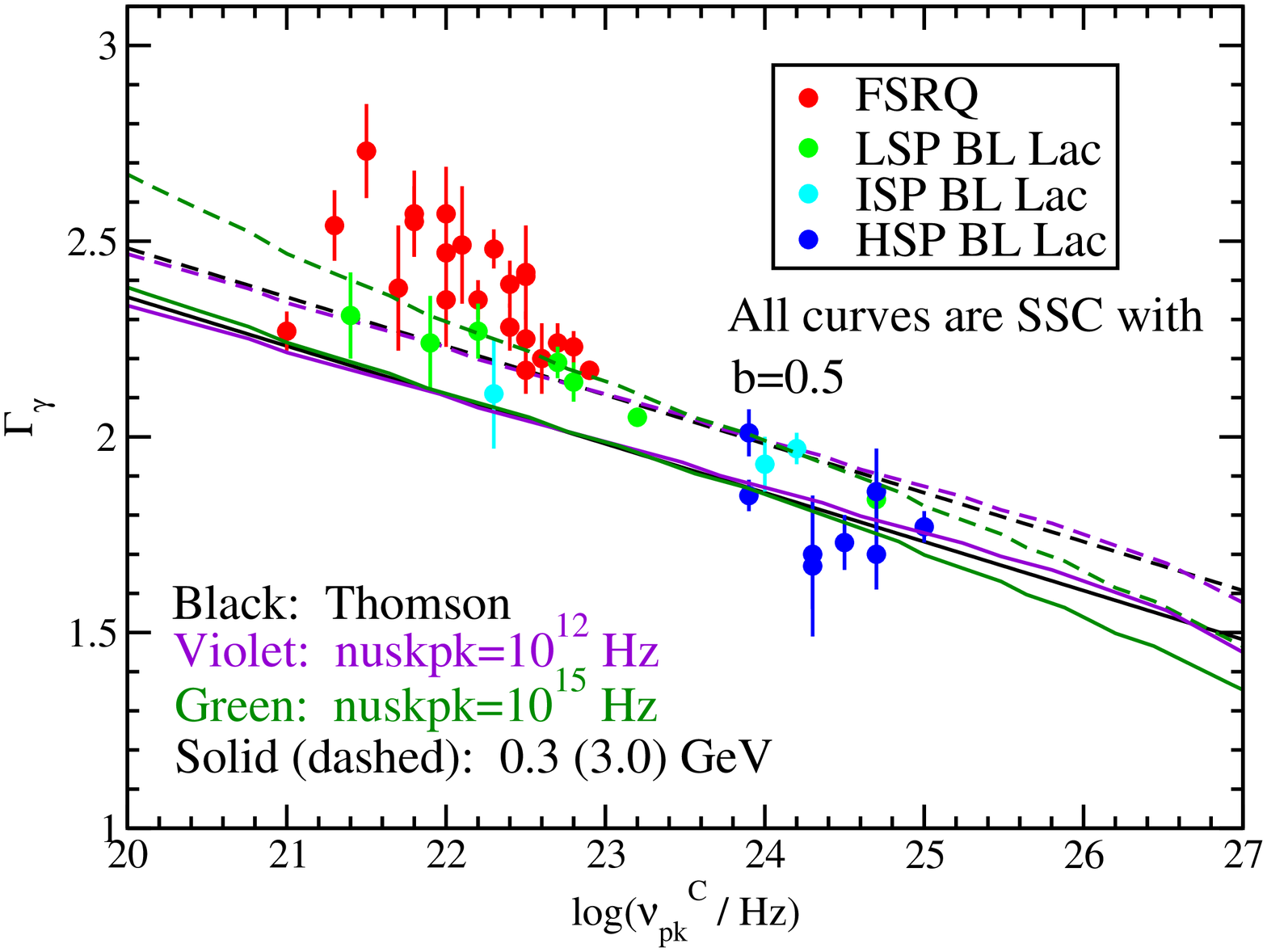}
 \caption{Data points show the  Fermi-LAT  $\gamma$-ray spectral index evaluated
in the range $0.1$ 1-- 100 GeV 
as a function of $\nu F_\nu$ peak Compton
frequency $\nu_{\rm C} = \nu_{pk}^{\rm C}$ of the blazar $\gamma$-ray 
SED \citep{SED}. In both panels, the range to calculate the Fermi-LAT 
spectral index is at 0.3 GeV and 3 GeV 
for the solid and dashed curves, respectively.  {\it (a), upper}: Equipartition 
Thomson-model EC predictions (black)
are shown along with numerical predictions evaluated for external 1000 K radiation
fields from a dusty torus (magneta curves) and from Ly$\alpha$ radiation (orange), 
using parameters of Fig.\ \ref{fig1} but with $b = 1/2$. 
{\it (b), lower}: Equipartition Thomson-model SSC predictions (black)
are shown along with numerical SSC predictions resulting from 
synchrotron emission with $\nu_s = 10^{12}$ Hz and $10^{15}$ Hz, as labeled.
 }
\label{fig5}
\end{center}
\end{figure}
%\vskip0.2in

\subsection{ Spectral index vs.\ peak Compton frequency }

Fig.\ \ref{fig5} shows data from multiwavelength spectral analysis \citep{SED} of 48 bright blazars in the
Fermi-LAT Bright AGN Sample \citep[LBAS;][]{LBAS}, separated into FSRQs, and low, intermediate, and high 
synchrotron-peaked (LSP, ISP, and HSP, respectively, defined by whether $\nu_s<10^{14}$ Hz,  $10^{14}<\nu_s$(Hz)$<10^{15}$ Hz, or
$\nu_s>10^{15}$ Hz) BL Lac objects.
The upper and lower panels gives predictions for the dependence of $\Gamma_\gamma$ on $\nu_{\rm C}$ for
the equipartition EC and SSC models. The Thomson-regime predictions, eqs.\ (\ref{Gg}) for EC processes and eq.\ (\ref{GgSSC}) for SSC processes, are plotted in black, depending on whether the $\gamma$-ray spectral index is measured at 0.3 GeV (solid curves)  or 3 GeV (dashed curves). The index is softer
when the $\gamma$-ray energy range used to determine the spectral index is larger, as noted above.

It is worth taking a moment to explain the deviations of the numerical curves from the Thomson-regime expressions.
Suppose the detector waveband $E_{GeV}\gg h\nu_{\rm C}$, corresponding to the left potions of the figures for $E_{GeV} = 0.3$ -- 3. 
Consider two $\gamma$-ray SEDs aligned at the same value of $\nu_{\rm C}$, one with strong Klein-Nishina effects and one in the Thomson regime. The SED with strong KN effects will be much softer at frequencies $\nu \gg \nu_{\rm C}$ by comparison with the one in the Thomson regime, causing the softer spectra when $h\nu_{\rm C}\ll E_{GeV}$, sometimes dramatically so, compared to SEDs formed by scattering in the Thomson regime.

At the other extreme $h\nu_{\rm C}\gg E_{GeV}$, corresponding to the right portions of the figures, the effects of strong KN losses is to harden the low-energy portion of the $\gamma$-ray SED compared to an SED formed by scattering in the Thomson regime (and with the same peak Compton frequency). Consequently, Klein-Nishina effects will produce harder spectra when the detector energy range is less than the peak Compton frequency compared to Thomson scattering.

%$\gamma_p^2 Q(\gamma_p, \Omega_p) = \delta_{\rm D}^4  \gamma_p^{\prime 2} Q^\prime(\gamma_p^\prime,\Omega_p^\prime )
%=  {\delta_{\rm D}^4 {\gamma_p^{\prime 2}} N^\prime(\gamma_p^\prime)\over 4\pi  t^\prime_{esc} }  $

%Because the standard blazar model for FSRQs has an EC origin of the $\gamma$ rays,
%the equipartition model explains the Fermi-LAT data.  
Fig.\ \ref{fig5} shows that an EC origin in either BLR or IR 
radiation is consistent with LSP FSRQ data,  but is
inconsistent with an SSC origin. 
A similar
 conclusion was reached earlier by examining the correlation of Compton 
dominance with core dominance 
in FSRQs and BL Lac objects \citep{mey12}.
At values of
$\nu_{\rm C} \gg 10^{23}$ Hz, or $E_{\rm C} \gg 1$  GeV, the sources are all ISP and HSP
BL Lac objects, and are compatible with either an SSC or EC origin of the emission, given the uncertainties
in $t_{var}$. In principle, however, an EC origin can be distinguished from an SSC origin 
by comparing the curvature of the $\gamma$-ray component with that of the synchrotron component.

%In the Thomson regime, furthermore, the change in spectral index, $\Delta\Gamma_\gamma$, determined for two values of $E_{GeV}$, or in two ranges of energies used to measure $\Gamma_\gamma$, is systematically larger for EC processes than for SSC processes as a consequence of the narrower SEDs in EC $\gamma$ rays compared to the SSC emission. Determining the systematic shifts in spectral index derived in different energy ranges using Fermi-LAT data can in principle decide if a population is radiating $\gamma$ rays from SSC or EC processes.

\section{Non-Equipartition Model for BL Lac Objects}

Blazars may be out of equipartition, though
extremely out-of-equipartition blazars would be less favored
because of the additional power required.
Up to now, we have assumed that the equipartition parameter $\zeta_e = 1$,
which minimizes jet power for a given synchrotron SED and variability time,
assuming small baryon-loading.  Modeling of
3C 279 with $\zeta_e = 1$ was possible in \citet{der14a}, though the very highest
energy $\gamma$ rays were only successfully fit by using long variability times
with $t_{var} \approx 10^5$ -- $10^6$ s, in which case the X-ray emission
was not well fit \citep[cf.][]{hay12}. Better fits were found in 
the modeling of 3C 454.3 by taking $\zeta_e$ between 0.6 and 3.5 \citep{cer13}, which 
has a minor effect on the spectral slope relation.\footnote{The fitting published
in \citet{cer13} lacked log-parabola $b$-dependent factors derived in reply 
to the referee of \citet{der14a}. Updated values have $\zeta_e\sim 1$ and $\zeta_s\gtrsim 0.2$.}

\begin{figure}[t]
\begin{center}
 \includegraphics[width=3.8in]{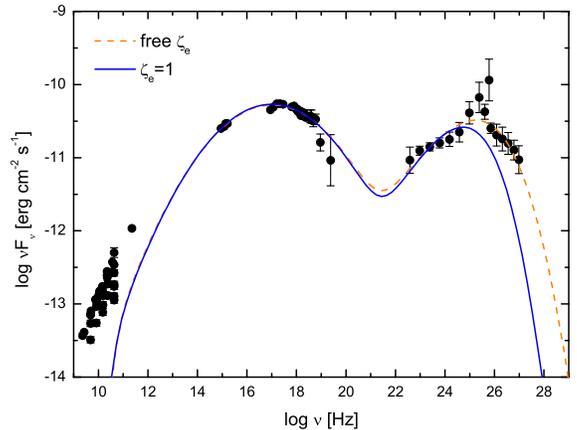}
 \caption{Best-fit models for Mrk 501 for fixed $\zeta_e =1$
and letting $\zeta_e$ vary.  Data from \citet{abd11MRK501}, with galactic
feature removed.}
\label{fig6}
\end{center}
\end{figure}

It is worth asking if an equipartition situation applies to BL Lac objects, which 
would be simpler than FSRQs by lacking significant external radiation fields.
We apply the near-equipartition log-parabola (NELP) modeling technique 
to the 15 March 2009 -- 1 Aug 2009 multiwavelength data of the 
HSP BL Lac object Mrk 501 \citep{abd11MRK501}.
The data in Fig.\ \ref{fig6} include OVRO radio observations, optical data, Swift UVOT and XRT data, 
GeV $\gamma$-ray data from Fermi-LAT,
and  VHE data from  MAGIC.
Parameter values are derived using the Markov Chain Monte Carlo (MCMC) technique of \cite{yan13}
for Mrk 421 and \cite{pyz14} for Mrk 501, and using the 3-parameter log-parabola electron
spectrum, eq.\ (\ref{g2Ng}).  The fit to the TeV data is always bad in the $\zeta_e=1$ case.
The fit with $\zeta_e$ allowed to vary is obviously far better.

\begin{figure}[t]
\begin{center}
 \includegraphics[width=3.5in]{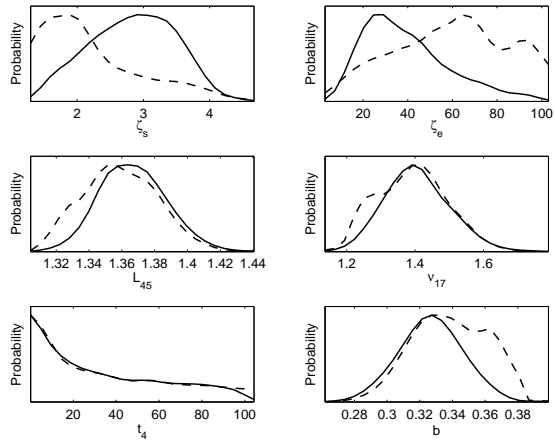}
 \caption{Distribution of parameter values for the varying $\zeta_e$ case. The dashed curves
represent the mean likelihoods of samples and the solid curves are the marginalized probabilities.  }
\label{fig7}
\end{center}
\end{figure}

The distribution of parameter values derived from the MCMC technique for the data of Mrk 501 is shown
in Fig.\ \ref{fig7}. The dashed curves are the mean likelihoods of samples 
and the solid curves are the marginalized probabilities.\footnote{See http://cosmologist.info/cosmomc/readme.html} 
In the fits, we run single chains and assume flat priors in the model parameter spaces. 
Since the MCMC code we used in this paper \citep{liu12,yuan11} is adapted
from COSMOMC, we refer the reader to \citet{lewis02} for a detailed explanation
of the code about sampling options, convergence
criteria, and statistical quantities. According to the results of 
\citet{yan13}, \citet{pyz14}, and \citet{zhou14}, the MCMC method is well suited to systematically 
investigate the high-dimensional model parameter spaces in fits to blazar SEDs.

\begin{figure}[t]
\begin{center}
 \includegraphics[width=3.5in]{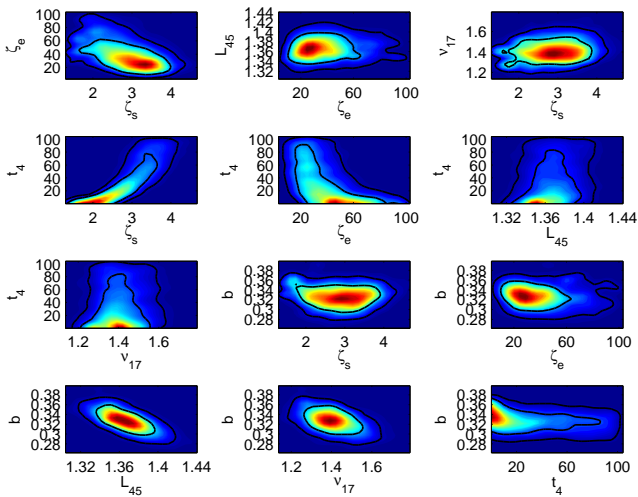}
 \caption{Two-dimensional probability contours of parameters.   }
\label{fig8}
\end{center}
\end{figure}

Pairs of values of $\zeta_s \cong 2, \zeta_e \cong 70$, and $\zeta_s \cong 3, \zeta_e \cong 30$,
from the fitting results shown in Fig.\ \ref{fig8} 
correspond to a change in index compared to an equipartition circumstance of
 $\Delta\Gamma_\gamma\cong -0.2b$ and $\Delta\Gamma_\gamma\cong -0.14b$, 
respectively. Even the large deviation from equipartition 
causes a spectral-index change $\lesssim 0.1$ unit for $b\cong 0.5$, and even less for $b\cong 1/3$.
The typical fluid-frame magnetic field derived from the fits has $B^\prime \approx 10$ mG. 
Synchrotron self-absorption is included in the fit.
Considerations about allowed jet power (see App.\ \ref{sec:AppB}) restrict the departure from equipartition further.  Thus
deviations from equipartition do not, on the basis of the Mrk 501 case, affect the spectral-index relation
significantly.

The inability of the numerical MCMC model to find a most favored value for $t_{var}$ 
may reflect limitations of the log-parabola EED used to model the Mrk 501 spectrum. Using a model 
joining a power-law at low electron energies with a log-parabola function at high electron
energies, \citet{pyz14} fit radio data down to $\approx$ GHz frequencies, and obtain
preferred variability times of $t_{var} \approx 5\times 10^5$ s. 

For given values of $L_{48}$ and $\nu_{14}$,
production of the highest energy $\gamma$-ray photons 
is assisted by going to an electron-dominated regime,
 where $\zeta_e\gg 1$ and $u_e^\prime \gg u^\prime_{B^\prime}$. 
The larger Lorentz factor electrons required to produce the 
same value of $\nu_s$ in a weaker magnetic field
can Compton scatter ambient photons to the highest energies.

\section{Discussion and Summary}
\label{sec:ds}

The nonthermal synchrotron paradigm pervades thinking in blazar physics, yet 
is incapable of explaining some of the most elementary facts, e.g., 
why synchrotron-radiating nonthermal electrons are apparently accelerated
so inefficiently.  Rather than reaching values of $\approx 100\Gamma$ MeV \citep[e.g.,][]{gfr83,jag96}, 
the peak synchrotron frequencies of FSRQs with $\nu_s \cong 10^{13}$ Hz are
$\approx 10^{10}$ times less than the highest energy synchrotron 
photon in the maximally efficient electron Fermi-acceleration scenario. 
Even the highest energy synchrotron photons
from HSP BL Lac objects rarely exceed $\approx 10$ -- 100 keV,
orders of magnitude below the radiation-reaction limit. It is crucial 
to understand the reason for the low peak synchrotron frequencies
(smaller, of course, than the maximum synchrotron frequency), and
how they relate to source luminosities and SEDs, which are
the basis of the  blazar sequence and blazar divide. 

\subsection{Near-equipartition, log-parabola (NELP) model} 

The astrophysics developed here may point a way to the   
solutions of these puzzles by first explaining the spectral-index diagrams. 
If the radiating electrons are near equipartition and approximately described by a log-parabola
EED because of the underlying acceleration and radiation physics, then 
the relationships between the $\gamma$-ray spectral index $\Gamma_\gamma$
and $\nu_s$ and $\nu_{\rm C}$ are precisely defined in the Thomson regime by 
functions of the form $\Gamma_\gamma = d - k\log\nu_{s({\rm C})}$, 
namely eqs.\ (\ref{GammagammaEC}) -- (\ref{GgSSC}). The slope is accurately
reproduced even when Klein-Nishina effects are important.
Moreover, the model inputs are all in principle 
observable from near-simultaneous multi-wavelength blazar campaigns: 
$L_{48}$, $\nu_s$, $\nu_{\rm C}$ and ${\cal A}_{\rm C}$ from spectral
observations,  $b$ and $\zeta_s$ from SED modeling, 
and $t_{var}$ from temporal analysis.
As shown here for Mrk 501, $\zeta_e$ and $\zeta_s$ can also be deduced from SED modeling, leaving only the
baryon-loading $\eta_{bl}$ as a major uncertainty, which affects the 
jet power (Appendix \ref{sec:AppB}). 

The near-equipartition approach using a 3-parameter log-parabola EED
 furthermore makes quantitative predictions about the  dependence of observables on  
$\Gamma_\gamma$ for statistical quantities of blazars, or for different states of a single blazar.
A specific example that can be performed with Fermi-LAT data is to determine 
$\gamma$-ray spectral indices of a large sample of blazars of specific types, e.g., LSP FSRQs and 
HSP BL Lac objects, in adjacent energy bands, giving the spectral curvature.
The curvature of the  $\gamma$-ray
SED is uniquely related to the  curvature of the synchrotron SED, depending on whether
the $\gamma$ rays have an SSC or EC origin. The difficulty of performing this test, of course,
is the requirement of quasi-simultaneous observations over a large energy range in order to provide 
a good characterization of the synchrotron SED peak and curvature.

For the synchrotron spectral-index diagram, our analysis
shows that  $k = k_{EC} = 3b/4$ for EC scattering in the Thomson regime, 
and $k = k_{SSC}  = 9b/16$ for synchrotron self-Compton (SSC) radiation. 
Numerical results show that this dependence is even valid when Klein-Nishina 
effects are important. Analysis of the combined sample of
FSRQs and BL Lac objects in the 3LAC \citep{3LAC}, $k_{3LAC} = 0.18\pm0.03$, 
implying curvatures of $b \cong 0.24$ if the emission arises
from EC processes, and $b \cong 0.32$ if the $\gamma$ rays are SSC.
However, it may not be correct to combine the two samples with different typical
values of $b$ in their populations.
Specific predictions 
for the slope of the $\Gamma_\gamma$ vs.\ $\nu_s$ behavior,
depending on whether the emission has an EC and SSC origin, 
should be studied for samples of blazars binned in ranges of 
$b$, because the two variables are correlated, with larger curvatures,
$b \approx 1$ for FSRQs, compared to $b\lesssim 0.5$ for BL Lac objects 
\citep{chen14}. Insofar as the SSC component seems less dominant in FSRQs
($\zeta_s \approx 0.2$) than in BL Lac objects ($\zeta_s \approx 1$), the effect
of this correlation on the spectral-index diagrams also has to be considered.

In principle, underlying correlations of $\Gamma_\gamma$ with  
 $t_{var}$ can be examined with the increasing number of simultaneous
multiwavelength blazar SEDs. Limitations of the log-parabola function
to describe the EED remains a central assumption that can be relaxed,
though not without  associated theoretical or numerical efforts.

\subsection{Departures from equipartition}

One of the uncertain parameters is the electron equipartition parameter $\zeta_e$, here
defined as the ratio of nonthermal electron and positron energy to 
magnetic-field energy throughout the volume of the radiating region.
Assuming $\zeta_e \cong 1, \zeta_s = 1$ gives the model predictions shown
in the Figs.\ \ref{fig1} -- \ref{fig4}. As shown in App.\ \ref{sec:AppB}, large departures from 
equipartition are not allowed if the absolute jet power is required to 
be less than the accretion power, which in turn is assumed to be bounded by the 
Eddington luminosity.\footnote{Counter-examples to this assumption are claimed
\citep{ghi14}. See App.\ \ref{sec:AppB}.} From the results of App.\ \ref{sec:AppB}, 
one possibility is that the SSC bolometric luminosity in the SEDs of 
large Compton-dominance FSRQ flaring events should
be small compared to the bolometric synchrotron luminosity (that is, $\zeta_s\ll 1$) 
for compatibility with sub-Eddington jet powers.

Spectral modeling of the FSRQs 3C 279 and 3C 454.3 is possible for $\zeta_e \cong 1$ 
\citep{cer13,der14a}.
For the BL Lac Mrk 501, a large departure from equipartition is 
required to get a good spectral fit, as we have shown, but even in this case, 
the effect from this out-of-equipartition condition on $\Gamma_\gamma$ is small. 
The deviations from equipartition giving 
the best fits to the SEDs of Mrk 501 show that
large $\zeta_e$, electron-particle--dominated fits (with
correspondingly weak magnetic fields) are favored to 
fit HSP BL Lacs extending into the TeV regime. 

\subsection{ Extensions of the log-parabola model}

{ The log-parabola function, eq.\ (\ref{g2Ng}), is motivated by second-order 
Fermi acceleration theory where MHD turbulence in the emitting fluid systematically accelerates
particles to form a curving EED (see Section 2). An equally compelling scenario combining
first- and second-order processes  considers a power-law distribution of particles
 injected downstream of a shock into a turbulent region where second-order processes
broaden the  distribution, so that the EED approximates the PLLP function, eq.\ (\ref{PLLP}). 

The full PLLP model has 5 parameters, but we have treated  in Figs.\ 1 -- 4 the important case
of an EED with a $-2$ number index extending to low energies without cutoff.
This EED makes a low-energy boundary 
to the spectral-index data near the Thomson value of $\Gamma_\gamma = 3/2$.  Remarkably,
this is as hard as the hardest Fermi-LAT blazar spectral indices measured.  
So if Compton scattering is responsible for the formation of the $\gamma$-ray SEDs of HSP blazars, 
as is undoubtedly true for the bulk of the radiation, then the PLLP model would give 
a simple explanation for the lack of blazars harder than $\Gamma_\gamma = 3/2$.
  
The situation is complicated, however, by the possibility that if the EEDs had low-energy
cutoffs rather than power-law extensions to low energies, then HSP blazars in the LAT band 
would tend to be dim and hard to  detect. So the apparent lack of blazars harder than $\Gamma_\gamma
= 3/2$ could be a selection effect rather than a limit imposed by the radiation physics. 
Searches in the Fermi-LAT $\gamma$-ray data for blazars harder than $\Gamma_\gamma
= 3/2$ would test whether a PLLP model is preferred over a LP model; the detection of such 
hard blazars would rule out the form of the PLLP considered here.

Except when $\nu_s\gg 10^{16}$ Hz, 
the existence of a low-energy power law in the EED makes 
only a small difference to the SSC predictions compared to the pure LP model.
Figs.\ 1 -- 4 show that the SSC predictions tend to be slightly softer than the data.
A low-energy cutoff in the EED could harden the SEDs, making it possible to attribute
an SSC origin to the $\gamma$-ray components to all HSP blazars. The discovery of 
blazars  harder than  $\Gamma_\gamma = 3/2$ in the Fermi-LAT energy range
would support this interpretation.
}

\subsection{Blazar types in the $\Gamma_\gamma$ vs.\ $\nu_s, \nu_{\rm C}$ plane }

We now ask why there are { essentially} no FSRQ blazars with $\nu_{s}\gtrsim 10^{14}$ Hz.
%%FSRQs are presumed to have BLR and IR radiation fields with 
%%energy densities $\gtrsim 10^{-3}$ and $\gtrsim 10^{-4}$ erg cm$^{-3}$, respectively.
The answer is likely to 
involve the dynamics of increasingly higher synchrotron-peaked 
near-equipartition jets
which, when finding themselves in an external radiation 
field, are subject to a radiation force opposite to the 
direction of motion that acts on the nonthermal electron 
population.  In the ideal one-zone model considered here,
there is no radiative drag from synchrotron and SSC processes, only from EC
processes \citep{tra11}.
The larger values of  $\delta_{\rm D}$ and $\gamma_{pk}^\prime$ 
for increasing $\nu_{s}$ implies a correspondingly larger
radiative drag when external radiation fields are present 
that would either slow the jet plasma down   \citep{gt10} or 
prevent it from reaching such large $\Gamma$ factors in the first place.

Rather than treating the jet dynamics, which is beyond the 
scope of the present enquiry,  we quote simple  analytic expressions 
for the synchrotron and external Compton SEDs from log-parabola
distribution, eqs.\ (\ref{eLsynesOmegas}) and (\ref{eLECesOmegas}),
which effectively answers the question
of how the Compton dominance ($\propto$ radiative drag)
grows with increasing $\nu_s$. 
From eq.\ (\ref{eLsynesOmegas}), to keep the apparent synchrotron 
luminosity constant requires 
$\delta_{\rm D}^4B^{\prime 2} \gamma_{pk}^{\prime 2} \sim $
 constant. Eq.\ (\ref{eLECesOmegas}) shows that the apparent
external Compton component grows $\propto \delta_{\rm D}^6 u_0 \gamma_{pk}^{\prime 2}$.
The ratio of the EC to synchrotron component is essentially the Compton dominance, 
which goes $\sim \delta_{\rm D}^2u_0/B^{\prime 2}\propto \nu_{14} L_{48}^{1/2}u_0$.
So the Compton dominance and radiation drag grow  $\propto\nu_{14}$, all other things
being equal. This confirms the behavior shown in Figs.\ \ref{fig1} -- \ref{fig4}, which 
deviates at large values of $\nu_{14}$ due to Klein-Nishina effects.
Depending on baryon loading, the jet could be quenched before escaping
the BLR, making an unusual, short flaring event. In such an unstable situation,
persistent emissions with large $\nu_{14}$ 
in dense external radiation environments might not be possible.

%This approach assumes a one-zone model fit to the broadband emission, so that 
%extended VHE jet emission \citep{bdf08,yzz12}, spine-sheath structure or decelerating jet emissions, 
%or VHE emission induced by UHECRs produced by the jet, are assumed to be unimportant.

\subsection{Maximum particle energy}

The near-equipartition log parabola model can be used to derive 
expressions for maximum escaping proton or ion energy $E_{max}$.
Starting with the \citet{hil84} condition in the form
$E_{\max} = Ze c B^\prime \delta_{\rm D}^2 t_{var}$
implies
\begin{equation}
E_{\max}({\rm eV}) = 1.4\times 10^{20}Z  L_{48}^{5/16}\;({t_4\over \nu_{14}})^{1/8} {f_1^{1/4} f_2^{1/8}\over \zeta_e^{1/4} \zeta_s^{1/16}f_0^{11/16}}\;,
\label{Emax}
\end{equation}
 using the  dependences given in Table \ref{table1} for the NELP model.
For a BL Lac object with $L_{48}\lesssim 0.01$, the only way to accelerate
ultra-high energy cosmic-ray
protons to $\gtrsim 10^{20}$ eV occurs when $\zeta_e\ll 1$, that is, in
a magnetically-dominated jet.
It is interesting to compare this expression with the formula
\begin{equation}
E_{max}({\rm eV}) = 2\times 10^{20} Z {\sqrt{\epsilon_B (L_{ph}/10^{48} {\rm ~erg~s}^{-1})/\epsilon_e}\over \Gamma/10 }\;
\label{Emax1}
\end{equation}
\citep{wax04,fg09,dr10}, 
which was also derived from the Hillas condition, 
 where $L_{ph}$ is the isotropic bolometric photon luminosity, and $\epsilon_B$ and $\epsilon_e$ are the 
fractions of jet power going into magnetic field and electrons, respectively.

An interesting feature of a combined lepto-hadronic blazar model using log-parabola functions for the 
particle distributions is that the synchrotron radiation-reaction limit for protons is $\approx 200\Gamma$ GeV, 
a factor $m_p/m_e$ greater than the electron limit. The evolution of the combined lepton synchrotron/SSC and 
proton synchrotron SEDs with $\zeta_e$ would favor a proton synchrotron component in the same large magnetization
($\zeta_e\ll 1$) regime where
the electrons are incapable of making high-energy radiation. 
%This also implies that a strongly magnetized
%flare is at a disadvantage to make the highest energy $\gamma$ rays from leptonic processes.

\subsection{Blazar sequence and blazar divide}

Multiwavelength data from any given blazar 
display a rich array of spectral and variability 
properties. The spectral properties of blazars  
in this  analysis are reduced to $\nu_s$, $L_{syn}$, $\nu_{\rm C}$, 
$\Gamma_\gamma$ and $b$, while the 
variability properties are reduced to $t_{var}$.

The spectral-index diagrams show robust correlations, 
which we explain as a consequence of relativistic blazar jets with
different powers and in different environments, within
which are  entrained relativistic electrons that can be 
described by log-parabola EEDs. By relating the synchrotron peak
frequency and synchrotron SED, which mirrors the EED,
to the spectral index of the $\gamma$-ray  SED 
formed through EC or SSC processes, 
the dependence of $\Gamma_\gamma = d - k\log \nu_s$ is 
easily derived in the Thomson regime. Moreover, the 
equipartition relations imply specific predictions
for underlying correlations.

The spectral-index diagrams are one side of a triangle
relating $\Gamma_\gamma$, $\nu_s$ (or $\nu_c$), and 
$L^{iso}$. The term $L^{iso}$
 can either be the apparent isotropic synchrotron, 
$\gamma$-ray, or total bolometric luminosity. 
The other two sides of the triangle are
 $L^{iso}$ vs.\ $\nu_s$ or $\nu_{\rm C}$, the blazar-sequence relations, and 
$\Gamma_\gamma$ vs.\ $L_\gamma$, the blazar-divide relation.

Our work illuminates one side of the triangle, namely $\Gamma_\gamma$ vs.\
$\nu_s$ or $\nu_{\rm C}$.
Regarding the blazar divide,
 suppose as a first approximation that the typical mass
of a supermassive black hole is $10^9 M_\odot$, then Fermi-LAT
data shows a significant change of spectral index at the Fermi 
divide of $L_\gamma \cong 10^{46}$ erg s$^{-1}$. 
If the apparent $\gamma$-ray luminosity is 10\% of the apparent
jet power, and the beaming correction is $\sim 100$, then
the divide is at $L/L_{\rm Edd} \cong 0.01$ \citep{gmt09}, and this would 
also represent the Eddington ratio below which the external radiation 
field energy density becomes small.

Extremely weak dependences, if any, are seen
in the $\Gamma_\gamma$ vs.\ $L_\gamma$ 
blazar-divide plots on either side of the divide.
Within blazar subpopulations \citep[see Fig.\ 39 in][]{2LAC},
the dependences of $\Gamma_\gamma$ on $L_\gamma$ are also weak.
Other than near the divide itself,
there is no clear dependence of $\Gamma_\gamma$ on blazar luminosity.
Indeed, any such dependence is predicted to be weak, as can be seen from 
eqs.\ (\ref{GammagammaEC}) and (\ref{GammagammaSSC}), which show 
that $\Gamma_\gamma\propto -b\log L_{syn}/8$ for EC processes, 
and $\Gamma_\gamma \propto b\log L_{syn}/32$ for SSC processes.

To explain the blazar sequence relating $L^{iso}$ and $\nu_s$ or $\nu_{\rm C}$
requires jet physics outside the scope of the present investigation. 
Rather than saying why blazars of a certain type can exist, however, 
we can suggest why blazars dominated by EC emission require low-synchrotron peaks. 
The presence
of any appreciable external radiation field would produce a Compton drag that 
 decelerates the bulk flow or prevents such a near-equipartition 
situation that would produce a synchrotron SED peaking at such large $\nu_s$ 
from forming.

\section{Summary}

To conclude, we have used an equipartition blazar modeling approach \citep{cer13,der14a} 
to explain the correlations of Fermi-LAT $\gamma$-ray number spectral index 
$\Gamma_\gamma$ with peak synchrotron frequency $\nu_s$
and peak Compton frequency $\nu_{\rm C}$. 
This approach assumes a one-zone model fit to the broadband emission, so that
emissions from, e.g.,
extended VHE jets \citep{bdf08,yzz12}, 
spine-sheath structures \citep{gtc05}, decelerating jets \citep{gk03}, 
or VHE emissions induced by UHECRs produced by the jet \citep{ess10,tmd13}, 
{ are assumed not to} affect the $\gamma$-ray spectral indices or peak frequencies.
Within this framework, the trends in the 
spectral-index diagrams are reproduced in a model with 
equipartition conditions and a log-parabola
electron distribution with $b \cong 1/2$. This conclusion 
holds even for out-of-equipartition conditions 
%provided they are 
limited by absolute jet power to be sub-Eddington.

{ The broadly distributed data in the spectral-index diagrams 
suggest that a better model comparison would 
consider a distribution of parameter values
to define a preferred model region in the $\Gamma_\gamma$ vs.\ 
$\nu_s$ and $\nu_{\rm C}$ diagrams. 
Such an approach depends on knowing whether the 
correlation of $b$ with $\nu_s$ \citep{chen14} is robust, if one is to 
sample from a distribution in $b$ values. 
Nevertheless, allowed regions in the 
spectral-index diagrams in Figs.\ 1 -- 5 are already defined  
by the heavy solid curves, depending on whether internal SSC or external 
EC BLR or EC IR processes dominate the formation of the $\gamma$-ray
SED. A distinct trend in 
the boundaries on the spectral index diagrams in Figs.\ 1 and 
3 are found for the PLLP model  that can be tested with 
Fermi-LAT analyses in different energy ranges.
%One set of parameters that can define a model prediction characterize The  target radiation fields for Compton scattering to $\gamma$-ray energies
%are given by the solid curves defining the EC BLR, EC IR, and SSC models . 

The $\Gamma_\gamma$ vs.\ $\nu_s$ boundaries defined by the dominance of 
internal SSC or external EC IR or EC BLR processes 
have, furthermore, a  very different shape when $b = 1$ 
(Fig.\ 2) compared to $b = 0.5$ (Fig.\ 1). This can be tested by 
subdividing the Fermi-LAT $\gamma$-ray spectral indices in different ranges of $b$.  
%The correlation of $b$ with peak synchrotron frequency $\nu_s$ \citep{chen14}, 
% if confirmed with more simultaneous blazar SED data, 
%would provide an important clue to the acceleration processes 
%in the blazars, and could be implemented to circumscribe different 
%boundaries in the $\Gamma_\gamma$ vs.\ $\nu_s$ plane defined by presence or 
% absence of external radiation fields. 
%A correlation of $b$ with $\nu_s$ could be implemented in the equipartition relations.
Whether the log-parabola function or the PLLP model, eq.\ (\ref{PLLP}) with a 
low-energy electron index $s=2$,  better approximates the EEDs can be tested by 
searching for Fermi-LAT sources with $\Gamma_\gamma < 1.5$.
}

The weak dependences
of $\Gamma_\gamma$ on changes in $L_{syn}$ found 
in eqs.\ (\ref{GammagammaEC}) and (\ref{GammagammaSSC})
are consistent with the weak dependences of 
$\gamma$-ray spectral index $\Gamma_\gamma$ on $L_\gamma$ 
on either side of the blazar divide. 
A physical explanation for  the change of the radiation environment 
of blazars at $\approx 0.01 L_{\rm Edd}$, though a 
reasonable model assumption, would make sense
of the blazar divide. 

This leaves open the blazar sequence relations, which
can ultimately only be understood from the physics occurring in the 
magnetospheres of the supermassive black holes powering
the blazars. Near-equipartition 
blazar synchrotron sources with $\nu_s\gg 10^{15}$ 
Hz would suffer increasingly strong radiation pressure in 
an environment with dense external radiation fields, which
could explain the absence of HSP FSRQs. Supermassive 
black-hole jets are most
luminous when their emissions are coolest, that is, 
when their peak synchrotron and Compton frequencies are lowest.
The near-equipartition log parabola blazar model 
provides a constrained system that explains
the spectral-index diagrams, and points to 
studies that could allow for a deeper understanding
of the blazar sequence and blazar divide.

\acknowledgements
The work of C.D.D.\ and J.D.F.\ is supported by the
Chief of Naval Research. We thank Dr.\ Matteo Cerruti
for discussions about spectral fitting, { and the 
anonymous referee for constructive questions and the recommendation
to consider the PLLP model}.

\appendix

\section{$\delta$-function Thomson-Regime SSC Derivation with Log-Parabola EED}

We derive the form of the $\nu L_\nu$ SED for the SSC component in the Thomson regime
assuming a log-parabola function of the EED and employing $\delta$-function approximations
for synchrotron and Thomson scattering. The comoving
Thomson-scattered synchrotron self-Compton
spectrum for isotropic distributions
of photons
and nonthermal relativistic electrons is given by
\begin{equation}
\e_1^\prime L^\prime_{SSC}(\e_1^\prime;\Omega^\prime) = {1\over 2} m_ec^3 \e_1^{\prime 2}
\int_1^\infty d\gamma^\prime \int_0^\infty d\ep\int_{-1}^1 d\mu^\prime (1-\mu^\prime )\; N^\prime_e(\gp )\; n_{ph}^\prime(\ep )\;{d\sigma(\bar\epsilon )\over d\e^\prime_1 }\;.
\label{esLSSC}
\end{equation}
Here, $\mu^\prime $ is the cosine of the angle between the directions of the interacting electron and photon, $\bar \e = \g\ep(1-\mu^\prime )$ is the invariant collision energy, and $d\sigma(\bar\e )/d\e^\prime_1$ is the differential scattering cross section.
We use the $\delta$-function Thomson scattering cross section $d\sigma(\bar\e )/d\e^\prime_1 = \sigma_{\rm T} \delta [\e^\prime_1 -\gamma^{\prime 2}\ep (1-\mu^\prime )]$ \citep[eq.\ (6.44);][]{dm09}.
From eq.\ (\ref{g2Ng}), $N_e^\prime(\gp ) = K^\prime y^{-2-b\log y}/ \gamma_{pk}^{\prime 2}$, and the photon spectral density
$n^\prime_{ph}(\ep ) = \ep L^\prime (\ep ) / 4\pi f_0 R_b^{\prime 2}\e^{\prime 2}m_ec^3$, where $f_0$ is a geometry factor, $R_b^\prime = c \delta_{\rm D} t_{var}$, and $t_{var}(1+z)$ is the measured variability time \citep{der14a}. For the synchrotron target photon spectrum,
$\ep L_{syn}^\prime (\ep ) = \e L_{syn}(\e )/\delta_{\rm D}^4 = \upsilon x^{1-b\log x}/\delta_{\rm D}^4$. Plugging these expressions into
eq.\ (\ref{esLSSC}), and using the $\delta$-function to solve the $\mu^\prime$ integral, we find
\begin{equation}
\e_1^\prime L^\prime_{SSC}(\e_1^\prime;\Omega^\prime) = {\sigma_{\rm T} \upsilon K^\prime \over \delta_{\rm D}^4 4\pi f_0 R_b^{\prime 2} \gamma_{pk}^{\prime 5} \epsilon_{pk}^{\prime 3}} \int_{1/\gamma_{pk}^\prime }^\infty dy \; y^{-6-b\log y}\;
\int_{x_\ell}^\infty dx \;x^{-6-b\log x}\;.
\label{esLSSC1}
\end{equation}
Here $x_\ell \equiv \sqrt{A}/y$, where $A = \e_1^\prime /2\gamma_{pk}^{\prime 2} \e_{pk}^\prime = \epsilon/\e_{pk,SSC}$.
The interior integral can be solved by noting, to good approximation, the logarithmic term is slowly varying compared to
the $x^{-6}$ term. The value of this integral is then $x_\ell^{-5-b\log x_\ell}/5$. After some manipulations, we obtain
\begin{equation}
\e L_{SSC}(\e;\Omega ) \cong{2\sigma_{\rm T}\upsilon K^\prime\gamma_{pk}^\prime\over 5\pi f_0 R_b^{\prime 2}}\;
\sqrt{{\pi \ln 10\over 2 b}}\; A^{{1\over 2} - {b\over 8}\log A}\;.
\label{eLSSC}
\end{equation}
Comparing with eqs.\ (\ref{Lsyne}) and (\ref{alphanu}) shows that the SSC spectral index is given by eq.\ (\ref{alphanu})
with $b$ replaced by $b/2$ and $\e_{pk}$ by $\e_{pk,SSC}$, leading to eq.\ (\ref{GammagammaSSC}).

%\section{The Spectral-Index Diagram at $\gamma$-ray frequencies}

\section{Jet Power in the Near-Equipartition Log-Parabola Model}
\label{sec:AppB}
We consider jet power with the addition of baryons and photons. 
The baryon-loading factor $\eta_{bl} \equiv u^\prime_{p/i}/u^\prime_{e}$, and $\zeta_e \equiv u_e^\prime/u^\prime_{B^\prime}$,
where $u^\prime_{p/i}$ is the fluid energy density of both thermal and nonthermal protons and ions, and $u^\prime_e$ is the 
nonthermal lepton energy density, including both electrons and positrons,
For convenience, the thermal electron and positron energy density is assumed small. The absolute
jet power for a two-sided jet is given by \citep{cf93,cg08,gt10}
\begin{equation}
L_{jet} = 2\pi r_b^{\prime 2} \beta \Gamma^2 c u^\prime_{B^\prime} [1+\zeta_e(1+\eta_{bl})] + L_{ph}\;,
\label{Ljet}
\end{equation}
where the absolute photon power $L_{ph}$ comprises synchrotron and SSC radiations, 
each assumed to be emitted isotropically in the 
jet frame, and EC radiations, with its comparatively narrower beaming \citep{der95}. 
The absolute photon powers depend on the observing angle $\theta$ through the 
Doppler factor $\delta_{\rm D}$ \citep{gt10,der12}, giving the absolute jet power in the NELP model:
\begin{equation}
L_{jet} ({\rm erg~s}^{-1}) = (1+N_\Gamma^2)^2\,
\{ 4.0\times 10^{44} \sqrt{f_1 \sqrt{ {f_2 t_4\over \nu_{14}} \sqrt{{L_{48}^5\over f_0\zeta_s} }  }  } 
\; [{1\over \sqrt{\zeta_e}}  + 
\sqrt{\zeta_e}(1+\eta_{bl}) ]+ {2 (L_{SSC}^{iso}+L_{syn}^{iso})\over 3\delta_{\rm D}^2} + {2 L_{EC}^{iso}\over 5\delta_{\rm D}^2} (1+N_\Gamma^2)^2
\}
\label{Ljet1} 
\end{equation}
where the observing angle $\theta \equiv N_\Gamma/\Gamma \ll 1$ and $\Gamma \gg 1$. In this expression,
$L_{syn}^{iso}$, $L_{SSC}^{iso}$, and $L_{EC}^{iso}$ are the measured apparent isotropic bolometric synchrotron, SSC, and EC luminosities, respectively. The other two terms in eq.\ (\ref{Ljet1}) correspond to the magnetic-field, $\propto 1/\sqrt{\zeta_e}$,
and the particle power, $\propto \sqrt{\zeta_e(1+\eta_{bl})}$. 
The additional factor of $(1+N^2_\Gamma)^2$ narrows the focus of the $\gamma$-ray beam.
Eq.\ (\ref{Ljet1}) can be rewritten as
\begin{equation}
L_{jet} ({\rm erg~s}^{-1}) = 4.0\times 10^{44}\,(1+N_\Gamma^2)^2\,{L_{48}^{5/8} f_1^{1/2}\over (f_0\zeta_s)^{1/8}}\;
({f_2 t_4\over \nu_{14}})^{1/4}\;\sqrt{(1+\eta_{bl})(1+\eta_{ph})}\,
%\,\sqrt{f_1 \sqrt{ {f_2 t_4\over \nu_{14}} \sqrt{{L_{48}^5\over f_0\zeta_s} }  }}
%{ [{1\over \sqrt{\zeta_e}}  + 
\large( {1\over  w}+  w \large)\;,
\label{Ljet2} 
\end{equation}
where 
\begin{equation} 
w \equiv \sqrt{\zeta_e(1+\eta_{bl})\over 1+\eta_{ph}}\;,
\label{what}
\end{equation} 
and the radiation loading
\begin{equation} 
\eta_{ph} \equiv {u^\prime_{rad}\over u^\prime_{B^\prime}} = 
 {5.4 f_0 \zeta_s} \;[1+\zeta_s + 0.6{\cal A}_{\rm EC}(1+N_\Gamma^2)^2]\;.
\label{Aph}
\end{equation} 
The isotropic bolometric photon luminosity 
$L_{ph}^{iso} = L_{syn}^{iso} + L_{EC}^{iso}+L_{SSC}^{iso} = \eta_{ph}L_{syn}^{iso}$, where
 the external Compton dominance ${\cal A}_{\rm EC} = L_{EC}^{iso}/L_{syn}^{iso}$.

%${\cal A}_{\rm C} = (L_{EC}^{iso}+L_{SSC}^{iso})/L_{syn}^{iso}$.

From eq.\ (\ref{Ljet2}), 
the minimum power condition is defined by the condition $w = 1$. When the baryon loading factor $\eta_{bl}\ll 1$
and the radiation loading $\eta_{ph}\ll 1$,
the minimum power condition is defined by $\zeta_e = 1$. If the baryon-loading is arbitrary, but $\eta_{ph}\ll 1$,
the minimum power condition is defined by $\zeta_e = 1/(1+\eta_{bl})$. 
When $\eta_{bl}\gg 1$, the minimum power
condition also corresponds to a highly magnetized jet (in terms of the electron energy density), with a larger jet luminosity by a factor $\sqrt{1+\eta_{bl}}$ at minimum jet power
compared to a pure electron/positron jet.  
When  $\eta_{bl}$ and $\eta_{ph}$ take arbitrary values,  the minimum power condition
is defined by $\zeta_e = (1+\eta_{ph})/(1+\eta_{bl})$, and the minimum jet power increases 
$\propto \sqrt{(1+\eta_{bl})(1+\eta_{ph})}$. 

Eq.\ (\ref{Ljet2}) gives the absolute minimum power to make the observed radiations from a blazar jet. For example, if the angular extent $\theta_j$ of the jet exceeds $1/\Gamma$, the power is increased by $\approx (\Gamma\theta_j)^2$. 
%For bright blazars detected by the Fermi-LAT well above threshold, we can expect $N_\Gamma \cong 1$ (whereas sources detected near the threshold sensitivity would more likely have $N_\Gamma < 1$). Thus  
When observing at $\theta\approx 1/\Gamma$, the minimum jet power, $L_{jet} \approx 3\times 10^{45} L_{48}^{5/8}$ erg s$^{-1}$
can only be increased by one to to orders of magnitude before exceeding $L_{\rm Edd} \approx 1.3\times 10^{47}M_9$ erg s$^{-1}$ for a $10^9 M_\odot$ black hole, unless one demands unusually high photon efficiencies. For HSP BL Lac objects, with $L_{syn}\lesssim 10^{46}$ erg s$^{-1}$, there is no great difficulty in satisfying the Eddington limit, even far from equipartition. However, these very same objects are believed to be accreting at a rate $\lesssim 0.01 L_{\rm EDD}$, so even in this case, large departures from equipartition cannot be tolerated. 

Based on Fermi-LAT data,  \citet{ghi14}  argue that the absolute jet power $P_{jet}$
%the absolute jet power $P_{jet} \approx 10P_{rad} \sim 20 L^{iso}_{\gamma}/\Gamma^2$, with bulk Lorentz factor $\Gamma$ of the radiating jet's %plasma outflow in the range $10\lesssim \Gamma \lesssim 15$ in the range 
is larger than the accretion-disk luminosity, which is approximated as $10$ times the BLR luminosity. They also approximate
$P_{jet} \approx 10P_{rad}$, with the absolute radiation power $P_{rad} \approx  k_fL^{iso}_{\gamma}/\Gamma^2$, where
the factor $k_f = 8/3$ for synchrotron/SSC processes and $k_f = 32/5$ for EC processes, 
and the bulk Lorentz factor $\Gamma$ of the radiating jet's 
plasma outflow is stated to be in the range $10\lesssim \Gamma \lesssim 15$.
%find instances where the jet power is much greater than the accretion luminosity. 
The underlying assumption  is that the observer is looking 
at $\theta_0 = 1/\Gamma$ to the jet axis. Inspection of the photon power
in eq.\ (\ref{Ljet2}) shows how uncertain this assumption is given how 
much brighter fluxes are along the jet axis compared to fluxes from  
sources at $\theta_0\approx 1/\Gamma$.
 Using the relation $1+N_\Gamma^2 =2\Gamma/\delta_{\rm D}$, 
the photon power in eq.\ (\ref{Ljet1}) for a 2-sided jet is
\begin{equation}
P_{rad} = (1+N_\Gamma^2)^4 \,{L_{syn}+L_{SSC}\over 6\Gamma^2} + (1+N_\Gamma^2)^6 \, {L_{EC}\over 10\Gamma^2}. 
\label{Prad}
\end{equation}  
When viewing down the jet axis, 
these powers are $\approx 16$ (synchrotron/SSC) and $\approx 64$ (for EC) times less than 
the values used in the expression for $P_{rad}$ by \citet{ghi14}. 
The most powerful $\gamma$-ray sources have the largest core dominances
and brightness temperatures \citep[e.g.,][however, see \citet{sav10}]{pus09,kov09,lfw14}, suggesting that these sources are 
also the ones viewed almost along the jet axis. { Indeed, \citet{jor05}, Fig.\ 25, find
no blazar with viewing angle $\theta_0 > 2/\Gamma$, whereas a large number of both BL Lacs and FSRQs have
$\theta_0 < 1/2\Gamma$. A severe overestimation of the radiation and therefore jet power is made by not taking
this effect into account.}

For very powerful FSRQs like 3C 454.3, which has $L_{48}\sim 1$, a curious feature arises. Great flares exceeding $L_{EC}^{iso} \gtrsim 10^{50}$ erg s$^{-1}$ with large Compton dominance  $\gtrsim 100$ can be allowed while maintaining absolute jet power $L_{jet} \lesssim L_{\rm Edd}$ only if $\theta\ll 1/\Gamma$ and $\zeta_s\ll 1$, that is, $L^{iso}_{syn} \gg L^{iso}_{SSC}$, implying a small SSC component relative to the synchrotron component. The effect of decreasing $\zeta_s$ is to increase $\delta_{\rm D}$ and narrow the Doppler cone, making the beaming factor even smaller, so that { extreme apparent EC $\gamma$-ray powers lead to absolute jet powers that are sub-Eddington}. 
%Another point is that when $\zeta_e(1+\eta_{bl}) < 1$, that is, a magnetically-dominated jet, electrons are not accelerated to such high energies as when the magnetic field is weak, so that
%the Compton components do not reach such high energies. Not only do highly magnetized jets allow hadronic emission components, 
%but they may demand it, because the electrons may not be sufficiently energetic to explain the VHE data. 
Spectral modeling to give the relative SSC and EC powers depends on X-ray observations, for example, Swift and NuSTAR, Fermi-LAT observations at GeV energies, and ground-based VHE air Cherenkov arrays.

%\pagebreak


\begin{thebibliography}{}

\bibitem[Abdo et al.(2009)]{LBAS} Abdo, A.~A., Ackermann,
M., Ajello, M., et al.\ 2009, \apj, 700, 597


\bibitem[Abdo et al.(2010a)]{1LAC} Abdo, A.~A., Ackermann,
M., Ajello, M., et al.\ 2010a, \apj, 715, 429

\bibitem[Abdo et al.(2010b)]{SED} Abdo, A.~A., Ackermann, M., Agudo, I., et al.\ 2010b, \apj, 716, 30
%The Spectral Energy Distribution of Fermi Bright Blazars

\bibitem[Abdo et al.(2010c)]{1FGL} Abdo, A.~A., Ackermann, 
M., Ajello, M., et al.\ 2010c, \apjs, 188, 405 


%\bibitem[Abdo et al.(2010c)]{abd10c} Abdo, A.~A., Ackermann,
%M., Ajello, M., et al.\ 2010c, \apj, 710, 1271
%Spectral Properties of Bright Fermi-Detected Blazars in the Gamma-Ray Band

\bibitem[Abdo et al.(2011)]{abd11MRK501} Abdo, A.~A., Ackermann, 
M., Ajello, M., et al.\ 2011, \apj, 727, 129 


\bibitem[Ackermann et al.(2010)]{ack10} Ackermann, M.,
Ajello, M., Baldini, L., et al.\ 2010, \apj, 721, 1383


\bibitem[Ackermann et al.(2011)]{2LAC} Ackermann, M.,
Ajello, M., Allafort, A., et al.\ 2011, \apj, 743, 171


\bibitem[Ackermann et al.(2015)]{3LAC} Ackermann, M.,
Ajello, M., Atwood, W., et al.\ 2015, submitted to \apj, arXiv:1501.06054


\bibitem[Aharonian et al.(2007)]{aha07PKS2155} Aharonian, F., 
Akhperjanian, A.~G., Bazer-Bachi, A.~R., et al.\ 2007, \apjl, 664, L71 

\bibitem[Ajello et al.(2014)]{aje14} Ajello, M., Romani,
R.~W., Gasparrini, D., et al.\ 2014, \apj, 780, 73

\bibitem[Albert et al.(2007)]{alb07Mrk501} Albert, J., Aliu, E., 
Anderhub, H., et al.\ 2007, \apj, 669, 862 

\bibitem[Aleksi{\'c} et al.(2011)]{PKS1222} Aleksi{\'c}, J., 
Antonelli, L.~A., Antoranz, P., et al.\ 2011, \apjl, 730, L8 


\bibitem[Becker et al.(2006)]{bld06} Becker, P.~A., Le, T.,
\& Dermer, C.~D.\ 2006, \apj, 647, 539

\bibitem[B{\l}a{\.z}ejowski et al.(2000)]{bla00}
B{\l}a{\.z}ejowski, M., Sikora, M., Moderski, R.,
\& Madejski, G.~M.\ 2000, \apj, 545, 107

\bibitem[Bloom \& Marscher(1996)]{bm96} Bloom, S.~D., \& Marscher, A.~P.\ 1996, \apj, 461, 657


\bibitem[Blumenthal \& Gould(1970)]{bg70}
Blumenthal, G.~R., \& Gould, R.~J.\ 1970, Reviews of Modern Physics, 42, 237

\bibitem[B{\"o}ttcher
\& Dermer(2002)]{bd02} B{\"o}ttcher, M., \& Dermer, C.~D.\ 2002, \apj, 564, 86

\bibitem[B{\"o}ttcher et al.(2008)]{bdf08} B{\"o}ttcher, M.,
Dermer, C.~D., \& Finke, J.~D.\ 2008, \apjl, 679, L9

\bibitem[B\"ottcher et al.(2012)]{bhk12} B\"ottcher, M.,
Harris, D.~E.,
\& Krawczynski, H., eds.\ 2012, Relativistic Jets from Active Galactic Nuclei (Berlin: Wiley)


\bibitem[Cavaliere \& D'Elia(2002)]{cd02} Cavaliere, A., \& D'Elia, V.\ 2002, \apj, 571, 226


\bibitem[Celotti \& Fabian(1993)]{cf93} Celotti, A., \& Fabian, A.~C.\ 1993, \mnras, 264, 228 
\bibitem[Celotti \& Ghisellini(2008)]{cg08} Celotti, A., \& Ghisellini, G.\ 2008, \mnras, 385, 283 


\bibitem[Cerruti et al.(2013)]{cer13} Cerruti, M., Dermer,
C.~D., Lott, B., Boisson, C., \& Zech, A.\ 2013, \apjl, 771, LL4

\bibitem[Chen(2014)]{chen14} Chen, L.\ 2014, \apj, 788, 179

\bibitem[Dermer(1995)]{der95} Dermer, C.~D.\ 1995, \apjl, 
446, L63 

\bibitem[Dermer et al.(2012)]{der12} Dermer, C.~D., Murase, K., \& Takami, H.\ 2012, \apj, 755, 147 

\bibitem[Dermer \& Razzaque(2010)]{dr10} Dermer, C.~D., \& Razzaque, S.\ 2010, \apj, 724, 1366 




\bibitem[Dermer et al.(2014a)]{der14a} Dermer, C.~D., Cerruti, M., Lott, B., Boisson, C., \& Zech, A.\ 2014a, \apj, 782, 82

\bibitem[Dermer et al.(2014b)]{dmi14} Dermer, C.~D., Murase, K., \& Inoue, Y.\ 2014b, Journal of High Energy Astrophysics, 3, 29

\bibitem[Dermer \& Schlickeiser(2002)]{ds02} Dermer, C.~D., \& Schlickeiser, R.\ 2002, \apj, 575, 667


\bibitem[Dermer \& Menon(2009)]{dm09} Dermer, C.~D., \& Menon, G.\ 2009, High Energy Radiation from Black Holes (Princeton University Press)

\bibitem[Dermer et al.(1992)]{dsm92} Dermer, C.~D., Schlickeiser, R., \& Mastichiadis, A.\ 1992, \aap, 256, L27


\bibitem[Essey et al.(2010)]{ess10} Essey, W., Kalashev, O.~E., Kusenko, A., \& Beacom, J.~F.\ 2010, Physical Review Letters, 104, 141102 

\bibitem[Farrar \& Gruzinov(2009)]{fg09} Farrar, G.~R., \& Gruzinov, A.\ 2009, \apj, 693, 329 


\bibitem[Finke(2013)]{fin13} Finke, J.~D.\ 2013, \apj, 763, 134

\bibitem[Fossati et al.(1998)]{fos98} Fossati, G., Maraschi, L., Celotti, A., Comastri, A., \& Ghisellini, G.\ 1998, \mnras, 299, 433

\bibitem[Fossati et al.(2008)]{fos08MRK421} Fossati, G., Buckley, J.~H., Bond, I.~H., et al.\ 2008, \apj, 677, 906 


\bibitem[Georganopoulos et al.(2001)]{gkm01} Georganopoulos, M., Kirk, J.~G., \& Mastichiadis, A.\ 2001, \apj, 561, 111 
\bibitem[Georganopoulos \& Kazanas(2003)]{gk03} Georganopoulos, M., \& Kazanas, D.\ 2003, \apjl, 594, L27 

\bibitem[Ghisellini et al.(2009)]{gmt09} Ghisellini, G., Maraschi, L., \& Tavecchio, F.\ 2009, \mnras, 396, L105

\bibitem[Ghisellini et al.(1998)]{ghi98} Ghisellini, G., Celotti, A., Fossati, G., Maraschi, L., \& Comastri, A.\ 1998, \mnras, 301, 451

\bibitem[Ghisellini
\& Tavecchio(2009)]{gt09} Ghisellini, G., \& Tavecchio, F.\ 2009, \mnras, 397, 985

\bibitem[Ghisellini \& Tavecchio(2010)]{gt10} Ghisellini, G., \& Tavecchio, F.\ 2010, \mnras, 409, L79 


\bibitem[Ghisellini et al.(2014)]{ghi14} Ghisellini, G., Tavecchio, F., Maraschi, L., Celotti, A.,  \& Sbarrato, T.\ 2014, \nat, 515, 376 

\bibitem[Ghisellini et al.(2005)]{gtc05} Ghisellini, G., Tavecchio, F., \& Chiaberge, M.\ 2005, \aap, 432, 401 


\bibitem[Giommi et al.(2013)]{gpp13} Giommi, P., Padovani,
P., \& Polenta, G.\ 2013, \mnras, 431, 1914

\bibitem[Giommi et al.(2012)]{gio12} Giommi, P., Padovani,
P., Polenta, G., et al.\ 2012, \mnras, 420, 2899


\bibitem[Guilbert et al.(1983)]{gfr83} Guilbert, P.~W., 
Fabian, A.~C., \& Rees, M.~J.\ 1983, \mnras, 205, 593 

\bibitem[Hayashida et al.(2012)]{hay12} Hayashida, M.,
Madejski, G.~M., Nalewajko, K., et al.\ 2012, \apj, 754, 114

\bibitem[Hillas(1984)]{hil84} Hillas, A.~M.\ 1984, \araa, 22, 425 


\bibitem[de Jager et al.(1996)]{jag96} de Jager, O.~C., Harding, A.~K., Michelson, P.~F., et al.\ 1996, \apj, 457, 253 


\bibitem[Jorstad et al.(2005)]{jor05} Jorstad, S.~G., 
Marscher, A.~P., Lister, M.~L., et al.\ 2005, \aj, 130, 1418 




\bibitem[Kovalev et al.(2009)]{kov09} Kovalev, Y.~Y., Aller, 
H.~D., Aller, M.~F., et al.\ 2009, \apjl, 696, L17 

\bibitem[Lewis \& Bridle(2002)]{lewis02} Lewis, A., \& Bridle, S.\ 2002, PhRvD, 66, 103511

\bibitem[Li et al.(2014)]{lfw14} Li, S.~H., Fan, J.~H., \& Wu, D.~X.\ 2014, Journal of Astrophysics and Astronomy, 35, 467 


\bibitem[Liu et al.(2012)]{liu12}Liu, J., Yuan, Q., Bi, X. J., Li, H., \& Zhang, X. M.\ 2012, PhRvD, 85, d3507

\bibitem[Maraschi et al.(1992)]{mgc92} Maraschi, L.,
Ghisellini, G., \& Celotti, A.\ 1992, \apjl, 397, L5

\bibitem[Marscher \& Gear(1985)]{mg85} Marscher, A.~P., \& Gear, W.~K.\ 1985, \apj, 298, 114

\bibitem[Massaro et 
al.(2004)]{mas04} Massaro, E., Perri, M., Giommi, P., \& Nesci, R.\ 2004, \aap, 413, 489 


\bibitem[Massaro et 
al.(2004a)]{mas04a} Massaro, E., Perri, M., Giommi, P., Nesci, R., \& Verrecchia, F.\ 2004a, \aap, 422, 103 

\bibitem[Massaro et al.(2006)]{mas06}
Massaro, E., Tramacere, A., Perri, M., Giommi, P., \& Tosti, G.\ 2006, \aap, 448, 861

\bibitem[Meyer et al.(2011)]{Meyer} Meyer, E.~T., Fossati,
G., Georganopoulos, M., \& Lister, M.~L.\ 2011, \apj, 740, 98

\bibitem[Meyer et al.(2012)]{mey12} Meyer, E.~T., Fossati, 
G., Georganopoulos, M., \& Lister, M.~L.\ 2012, \apjl, 752, LL4 

\bibitem[Paggi et
al.(2009)]{pag09} Paggi, A., Massaro, F., Vittorini, V., et al.\ 2009, \aap, 504, 821

\bibitem[Peng et al.(2014)]{pyz14} Peng, Y., Yan, D.,
\& Zhang, L.\ 2014, \mnras, 442, 2357


\bibitem[Pushkarev et 
al.(2009)]{pus09} Pushkarev, A.~B., Kovalev, Y.~Y., Lister, M.~L., \& Savolainen, T.\ 2009, \aap, 507, L33 

\bibitem[Sambruna et al.(1996)]{smu96} Sambruna, R.~M.,
Maraschi, L., \& Urry, C.~M.\ 1996, \apj, 463, 444

\bibitem[Savolainen et 
al.(2010)]{sav10} Savolainen, T., Homan, D.~C., Hovatta, T., et al.\ 2010, \aap, 512, AA24 


\bibitem[Shaw et al.(2013)]{sha13} Shaw, M.~S., Romani,
R.~W., Cotter, G., et al.\ 2013, \apj, 764, 135

\bibitem[Sikora et al.(1994)]{sbr94} Sikora, M., Begelman,
M.~C., \& Rees, M.~J.\ 1994, \apj, 421, 153

\bibitem[Sikora et al.(2009)]{sik09} Sikora, M., Stawarz,
{\L}., Moderski, R., Nalewajko, K., \& Madejski, G.~M.\ 2009, \apj, 704, 38

\bibitem[Stawarz
\& Petrosian(2008)]{sp08} Stawarz, {\L}., \& Petrosian, V.\ 2008, \apj, 681, 1725


\bibitem[Takami et al.(2013)]{tmd13} Takami, H., Murase, K., 
\& Dermer, C.~D.\ 2013, \apjl, 771, LL32 


\bibitem[Tavecchio
\& Ghisellini(2008)]{tg08} Tavecchio, F., \& Ghisellini, G.\ 2008, \mnras, 386, 945

\bibitem[Tramacere et 
al.(2007)]{tra07} Tramacere, A., Massaro, F., \& Cavaliere, A.\ 2007, \aap, 466, 521 


\bibitem[Tramacere et  al.(2009)]{tra09} Tramacere, A., Giommi, P., Perri, M., Verrecchia, F., \& Tosti, G.\ 2009, \aap, 501, 879 


\bibitem[Tramacere et al.(2011)]{tra11} Tramacere, A.,
Massaro, E., \& Taylor, A.~M.\ 2011, \apj, 739, 66


\bibitem[Waxman(2004)]{wax04} Waxman, E.\ 2004, New Journal of Physics, 6, 140 

\bibitem[Yan et al.(2012)]{yzz12} Yan, D., Zeng, H.,
\& Zhang, L.\ 2012, \mnras, 424, 2173

\bibitem[Yan et al.(2013)]{yan13} Yan, D., Zhang, L., Yuan,
Q., Fan, Z., \& Zeng, H.\ 2013, \apj, 765, 122

\bibitem[Yuan et al.(2011)]{yuan11}Yuan, Q., Liu, S., Fan, Z.,  Bi, X., \& Fryer, C.\ 2011, \apj, 735, 120

\bibitem[Zhou et al.(2014)]{zhou14}Zhou, Y., Yan, D., Dai, B., \& Zhang, L.\ 2014, PASJ, 66, 12


\end{thebibliography}
\end{document}